\documentclass{article}

\usepackage{algorithm}
\usepackage[margin=30mm,includehead,includefoot]{geometry}
\usepackage[noend]{algpseudocode}
\usepackage{amsmath,amssymb,amsthm,amscd,color,comment}
\usepackage{autobreak}
\usepackage{color}
\usepackage{xcolor}
\usepackage{url}
\usepackage{blkarray}
\usepackage{jheppub}
\usepackage{empheq}
\usepackage{graphicx}
\usepackage{booktabs, multirow}
\usepackage{subcaption}
\usepackage{bm}
\usepackage{enumerate}
\usepackage{comment}
\usepackage{parskip}
\usepackage{bbm}
\usepackage{slashed}
\usepackage{physics}
\usepackage{feynmp}
\definecolor{green1}{HTML}{3D792A}
\definecolor{cyan1}{HTML}{37cdaa}
\definecolor{blue1}{HTML}{5d7ac4}
\definecolor{red1}{HTML}{d0482a}
\definecolor{purple1}{HTML}{845ea8}
\definecolor{orange1}{HTML}{e07229}

\newcommand{\bega}{\begin{gathered}}
\newcommand{\ega}{\end{gathered}}

\title{Feynman Integral Reductions by Intersection Theory with Orthogonal Bases and Closed Formulae}

\author[a,b,c]{Giulio Crisanti,}
\author[a,b,d]{Sid Smith}

\affiliation[a]{\unipd}
\affiliation[b]{\pdinfn}
\affiliation[c]{\maxplanck}
\affiliation[d]{\higgs}

\preprint{MPP-2024-104}

\newcommand{\unipd}{Dipartimento di Fisica e Astronomia, Universit\`a degli Studi di Padova,
Via Marzolo 8, I-35131 Padova, Italy}

\newcommand{\pdinfn}{INFN, Sezione di Padova,
Via Marzolo 8, I-35131 Padova, Italy}

\newcommand{\higgs}{Higgs Centre for Theoretical Physics, University of Edinburgh, James Clerk Maxwell Building,Peter Guthrie Tait Road, Edinburgh, EH9 3FD, United Kingdom}

\newcommand{\maxplanck}{Max–Planck–Institut für Physik, Werner–Heisenberg–Institut, D–85748 Garching bei München, Germany}

\vspace{4cm}

\emailAdd{giulioeugenio.crisanti@phd.unipd.it}
\emailAdd{sid.smith@ed.ac.uk}

\abstract{
We present a prescription for choosing orthogonal bases of differential $n$-forms belonging to quadratic twisted period integrals, with respect to the intersection number inner product. To evaluate these inner products, we additionally propose a new closed formula for intersection numbers beyond $\dd \log$ forms. These findings allow us to systematically construct orthonormal bases between twisted period integrals of this type. In the context of Feynman integrals, this represents all diagrams at one-loop.
}

\allowdisplaybreaks

\begin{document}
\addtocontents{toc}{\protect\setcounter{tocdepth}{2}}
\begin{fmffile}{feyndiagrams}

\maketitle

\section{Introduction}

It has recently become clear that the study and computation of \textit{twisted period integrals} \cite{GOTO_2013,goto2013twisted,goto2014monodromy} is of the utmost importance to fundamental theories in many areas of physics. Generally speaking, these integrals can be viewed as a pairing between a differential form (cocycle) and a cycle. The pairing is simply the integration of the form over the contour, but with the addition of a multivalued function in the integrand known as the \textit{twist}. Integrals of this form belong to a vector space, which is isomorphic to a \textit{twisted de Rham cohomology group}. The (cohomology) \textit{intersection number} \cite{Matsumoto_1994,Cho_Matsumoto_1995,Matsumoto1998-2,ojm/1200788347,Mimachi_2003,Mimachi:2004ez,matsubaraheo2020computing,goto2020homology,matsubaraheo2021algorithm,matsubaraheo2022localization,mizerascatt,Mastrolia_2019,Frellesvig_2019} between differential $n$-forms is an elementary quantity that defines a pairing between forms of this group.

In the context of scattering amplitudes, Feynman integrals admit a vector space structure \cite{Mastrolia_2019,Frellesvig_2019,PhysRevLett.123.201602,Smirnov:2010hn,Lee_2013}, and can be viewed as twisted period integrals by transforming them to a parametric form such as Baikov or Lee-Pomeransky representation \cite{Baikov:1996iu,Lee_2013}. Thus, through the usage of intersection numbers, interpreted as inner products, any Feynman integral can be decomposed onto a master integral basis \cite{Mastrolia_2019,PhysRevLett.123.201602,Frellesvig_2019,Frellesvig_2021}. Therefore, integral family reductions and the construction of differential equations \cite{KOTIKOV1991158,GhermannRemiddi2000,Henn2013} can be performed in a completely complementary way to Integration by Parts (IBP) methods \cite{Chetyrkin:1981qh,Laporta:2000dsw}. %

In this formalism, reductions to master integrals are given by combinations of intersection numbers, given by the \textit{master decomposition formula} \cite{Mastrolia_2019}. Any reduction can be divided into two main parts. The first is the computation of the so called $\mathbf{C}$-matrix, or \textit{metric}, which is the matrix of intersection numbers between a chosen basis and its dual counterpart. The second stage is the computation of projections, namely the intersection numbers between any target integral, and the basis on which one wants to reduce it. 

Beyond Feynman integrals, many other problems admit a twisted period integral structure, and can be studied with intersection theory. Recently, canonical bases for differential equations \cite{Chen_2021,Chen_2022,Chen_2024}, lattice correlation functions \cite{Gasparotto_2023,Gasparotto_2023_2}, quantum mechanics \cite{cacciatori2022intersection}, cosmological correlators \cite{de2023cosmology}, Fourier integrals \cite{brunello2023fourier}, general relativity amplitudes \cite{brunello2024improved,frellesvig2024general}, string amplitudes \cite{Bhardwaj:2023vvm}, as well as much more \cite{Mizera_2020,Weinzierl_2020,Chestnov_2022,Giroux_2023,jiang2023recursive}, have all been tackled using intersection theory. 

Intersection numbers have also been studied in their own right, with many different strategies developed for their computation over the last few years \cite{mizerascatt,matsumoto2019relative,Weinzierl_2021,Caron_Huot_2021,Caron_Huot_2022,Fontana_2023,Chestnov_2022,Chestnov_2023,brunello2023intersection}. Recent developments include the removal of \textit{analytic regulators} in intersection computations for Feynman integrals. This was done through the introduction of \textit{relative twisted} intersection numbers \cite{matsumoto2019relative,Caron_Huot_2021,Caron_Huot_2022}, as well as a novel independent improvement to the regularisation approach presented in \cite{Fontana_2023}. These two strategies were shown to be connected in \cite{brunello2023intersection}.

Aside from computational algorithms, some properties of intersection numbers are well understood, particularly for $\mathrm{d}\log$ forms. In the case of hyperplanar arrangements \cite{ojm/1200788347,Mizera_2020_b}, but also beyond \cite{mizerascatt}, closed formulae are known for intersection numbers of this kind.

In this paper, we explore new properties of intersection numbers, specifically for twists originating from \textit{quadratic} multivariate polynomials. Given a Laporta basis, we provide an algorithm for choosing a dual basis that produces a diagonal $\mathbf{C}$-matrix, for relative intersection numbers. For $1$-form intersection numbers, such a procedure already exists \cite{Frellesvig_2019}, and similar results have been found for canonical bases \cite{Caron_Huot_2021}. In this paper however we focus our attention on Feynman integrals in Baikov representation. We observe that our procedure works best when the coefficients of the quadratic polynomial are ``generic". In practice, this requirement translates to one-loop Feynman diagrams with massive propagators. For massless diagrams our prescription also often works, with some minor modifications. Beyond one-loop, we also test our results for a simple two loop diagram, and find success in producing a block diagonal $\mathbf{C}$-matrix.

Our second result is a new, closed formula for intersection numbers with quadratic twists. The expression is remarkably simple, is the first of its kind beyond $\dd \log$ forms, and provides evidence of deep mathematical structures for objects of this kind. The main building blocks of the expression are two distinct \textit{multivariate discriminants}, expressed as determinants of Hessian matrices. The appearance of repeated univariate discriminants had previously been noticed in \cite{Caron_Huot_2021,Caron_Huot_2022}, and our expression connects to this but considers a different class of intersection numbers.

Combined with the diagonal prescription above, this result makes the computation of the $\mathbf{C}$-matrix extremely efficient for one-loop Feynman integrals, by requiring no explicit algorithmic computations of intersection numbers. Interpreted differently, these two findings combined allow us to construct not just diagonal, but identity $\mathbf{C}$-matrices.

The paper is structured as follows: In Section \ref{sec:review} we will give a brief review of intersection theory, explaining how intersection numbers are computed, and how they can be used to compute integral reductions. We also review relative intersection numbers, as well as outline how one-loop Feynman Integrals can be seen as twisted integrals from the Baikov representation. In Section \ref{sec:diagonal_basis_prescription} we describe the algorithm to choose orthogonal bases of differential forms, as well as provide a few initial examples of its usage. In Section \ref{sec:closed_formula} we present the closed form formula for intersection numbers, and show how this can be applied to compute every element in a diagonal $\mathbf{C}$-matrix. In Section \ref{sec:examples} we provide an extensive list of examples where we verify the results of Sections \ref{sec:diagonal_basis_prescription} and \ref{sec:closed_formula} to be working correctly. Finally, we provide concluding remarks and discuss possible extensions to this work in Section \ref{sec:conclusions}.

During our research the following software was used: \textsc{Mathematica}, \textsc{LiteRed} \cite{lee2012presenting,Lee_2014}, \textsc{FiniteFlow} \cite{Peraro_2019}, \textsc{JaxoDraw} \cite{Binosi_2009,VERMASEREN199445}.

\section{Review of Intersection Theory \label{sec:review}}

The central object of study of this work is the computation of intersection numbers for twisted integrals, with the ultimate goal of integral reductions onto bases. To this end in this section we provide a review of twisted integrals, intersection theory, and how intersection numbers can be used to compute IBP-like reductions for Feynman integrals.

\subsection{Twisted Cohomology}

We are concerned with \textit{twisted period integrals} of the form
\begin{equation}
    I = \int_{C_{R}}u(\mathbf{z})\ \varphi_{L}(\mathbf{z}) =: \bra{\varphi_{L}}C_{R}]\,.
\end{equation}
In this expression, $u(\mathbf{z})$ is a multi-valued function called the \textit{twist}, which takes the form
\begin{equation}
    u(\mathbf{z})=\prod_i b(\mathbf{z})_i^{\gamma_i}\,,
\end{equation}
where each $b_i(\mathbf{z})$ is a polynomial in $n$ variables $\mathbf{z}=\{z_1\cdots z_n\}$, and $\gamma_i$ is a generic exponent. $\varphi_{L}$ is a meromorphic $n$-form $\varphi_{L}(\mathbf{z})=\hat{\varphi}_{L}(\mathbf{z})\dd^{n}\mathbf{z}\equiv\hat{\varphi}_{L}(\mathbf{z})\dd z_{1}\wedge\dots\wedge\dd z_{n}$. We impose that the integration contour $C_{R}$ is defined such that $u$ vanishes on the boundary $u(\partial C_{R})=0$. This gives rise to integration-by-parts (IBP) identities of the form
\begin{equation}
    0 = \int_{C_{R}}\dd(u\,\xi_{L}) = \int_{C_{R}}u\,\grad_{\omega}\,\xi_{L}\,,
\end{equation}
where $\xi_{L}$ is some arbitrary differential $(n-1)$-form, and we define the covariant derivative $\grad_{\omega}$ as
\begin{equation}
    \omega:=\dd\log{u} = \sum_{i=1}^{n}\hat{\omega}_{i}dz_{i}\,, \quad \hat{\omega}_{i} = \pdv{\log{u}}{z_{i}}\,, \quad \grad_{\omega} = \dd+\omega\wedge\,.
\end{equation}
This allows us to define an equivalence relation between differential $n$-forms inside twisted integrals $\bra{\varphi_{L}}\sim\bra{\varphi_{L}+\grad_{\omega}\xi_{L}}$. The \textit{twisted cohomology group} $H_{\omega}^{n}$ is defined as the vector space of closed $n$-forms modulo this equivalence relation. This space identifies integrands that give the same $I$, upon integration with $C_{R}$. This vector space has finite dimension $\nu:=\text{dim}H_{\omega}^{n}$. We denote elements of this group $\bra{\varphi_{L}}\in H_{\omega}^{n}$.

There additionally exists a dual vector space, composed of \textit{dual integrals}, defined as
\begin{equation}
    \tilde{I} = \int_{C_{L}}u^{-1}\varphi_{R} = [C_{L}\ket{\varphi_{R}}\,.
\end{equation}
Its respective dual twisted cohomologoy group is given by $H_{-\omega}^{n}$, and we denote elements of this group by $\ket{\varphi_{R}}\in H_{-\omega}^{n}$.

The dimension $\nu$ of the twisted cohomology group counts the number of independent integrals that can be built from $n$-forms satisfying the conditions given above. For the case of Feynman integrals, this corresponds to the number of MIs. The dimension of the group is given directly by \cite{Lee_2013,Frellesvig_2021}
\begin{equation}
     \nu = \text{dim}\ H_{\pm\omega}^{n} = \{\#\text{ of solutions to }\omega=0\}\,.
\end{equation}

\subsection{Basis of Master Integrals}

Given a twist $u$ one can formally define a set of bases that span the (dual) cohomology group 
\begin{equation}
    \bra{e_{i}}\in H_{\omega}^n\,,\qquad\ket{h_{i}}\in H_{-\omega}^n\,,\qquad i=1,\dots,\nu\,.
\end{equation}
Any arbitrary $n$-form $\bra{\varphi_{L}}$ can then be decomposed in terms of this basis schematically as
\begin{equation}
    \bra{\varphi_{L}} = \sum_{i=1}^{\nu}c_{i}\bra{e_{i}}\,.
\end{equation}
This is equivalent to decomposing our original integral $I=\bra{\varphi_L}C_{R}]$ in terms of a set of Master Integrals $J_{i}=\bra{e_{i}}C_{R}]$:
\begin{equation}
    I = \bra{\varphi_{L}}C_{R}] = \sum_{i=1}^{\nu}c_{i}\bra{e_{i}}C_{R}] = \sum_{i=1}^{\nu}c_{i}J_{i}\,.
\end{equation}
The coefficients $c_i$ of this decomposition can be found by using the \textit{master decomposition formula} \cite{Mastrolia_2019}
\begin{equation}\label{eq:masterdecomp}
    c_{i} = \sum_{j=1}^{\nu}\braket{\varphi_{L}}{h_{j}}(\mathbf{C}^{-1})_{ji}\,,
\end{equation}
where we define
\begin{equation}
    \mathbf{C}_{ij} = \braket{e_{i}}{h_{j}}\,,
\end{equation}
which will be denoted as the $\mathbf{C}$-matrix, or metric in the rest of this work. Equation (\ref{eq:masterdecomp}) assumes that there exists an inner product between forms and dual forms, $\braket{\bullet}{\bullet}$. In the context of twisted period integrals, this object exists, is well defined, and is denoted as the \textit{intersection number} \cite{Cho_Matsumoto_1995,Matsumoto1998-2,Mimachi_2003,Mimachi:2004ez,ojm/1200788347,matsubaraheo2020computing,goto2020homology,matsubaraheo2021algorithm,matsubaraheo2022localization,mizerascatt,Mastrolia_2019,Frellesvig_2019}.

\subsection{Intersection Numbers}

The intersection number is an inner product between $n$-forms of a twisted cohomology group and its dual counterpart. Formally, it is defined as an integral over these two forms:
\begin{equation}
    \braket{\varphi_{L}}{\varphi_{R}} := \frac{1}{2\pi i}\int_{X}\iota(\varphi_{L})\wedge\varphi_{R}\,.
\end{equation}
Here $\iota$ defines a regularisation operator, which is necessary for the computation to be well defined. 

\subsubsection{Univariate Intersection Numbers}

Specifying for the moment to the case $n=1$, $\iota$ is defined as
\begin{equation}
    \iota(\varphi_{L}) := \varphi_{L}-\grad_{\omega}(h\psi_{L})\,.
\end{equation}
Here, $h$ is a Heaviside function
\begin{equation}
    h := \sum_{p\in\mathcal{P}({\omega})}\left(\theta(\epsilon-\abs{z-z_{i}})\right), \quad \mathcal{P}({\omega}) := \text{poles of $\omega$}\,,
\end{equation}
and the function $\psi_L$ is the solution to the differential equation:
\begin{equation}
    \grad_{\omega}\psi_{L}=\varphi_{L}\,.
\end{equation}
The space of integration is $X=\mathbb{C}\mathrm{P}\setminus\mathcal{P}({\omega})$.
After some algebra\footnote{See \cite{ojm/1200788347,mizerascatt,Mizera_2020_b,Gasparotto:2023cdl,weinzierl2022feynman} for derivations.}, the univariate intersection number localises to a straightforward residue computation:
\begin{equation}
    \braket{\varphi_{L}}{\varphi_{R}} = \sum_{p\in\mathcal{P}({\omega})}\text{Res}_{z=p}(\psi_{L}\varphi_{R})\,.
\end{equation}
For this procedure to be well defined, it is essential that the poles in $\varphi_L$ and $\varphi_R$ are \textit{regulated} by $u$, which we define to be the condition
\begin{equation}\label{eq:regulatedcondition}
    \mathcal{P}(\varphi_{L/R})\in\mathcal{P}({\omega})\,.
\end{equation}
 Many cases of physical interest involve twisted integrals that are not regulated. In Section (\ref{subsec:relcohom}) we discuss the \textit{relative cohomology} approach to overcoming this problem.

\subsubsection{Multivariate intersection numbers}

For the case $n>1$, there are several strategies one can adopt for intersection computations \cite{Frellesvig_2021,PhysRevLett.123.201602,Chestnov_2023,ojm/1200788347,Frellesvig_2019,Chestnov_2022}. For this work, we will use the \textit{recursive} or \textit{fibration} based approach, which we review now. For more details on this algorithm, see \cite{PhysRevLett.123.201602,Mizera_2020_b,Frellesvig_2021} and \cite{Gasparotto:2023cdl,weinzierl2022feynman} for reviews. 

The recursive algorithm builds $n$-form intersection numbers one variable at a time: for concreteness, we order them  as $z_{n},\dots,z_{1}$. We can consider the $m$-th ``layer" of the iterative process as treating the variables $z_{n},\dots,z_{m+1}$ as ``external", and the variables $z_{m},\dots,z_{1}$ as ``internal" integration variables. The $m$-forms generated by these internal variables belong to their own twisted cohomology group and thus also form a vector space. For all $1\leq m\leq n$, we denote the dimension of this inner space as $\nu_{m}$, and a basis of (dual) forms as $\bra*{e_{i}^{(\mathbf{m})}}$ and $\ket*{h_{i}^{(\mathbf{m})}}$ respectively\footnote{A systematic algorithm for choosing a basis of forms and duals is presented in Section \ref{subsec:standard_prescription}.}.

Let us assume all $(m-1)$-form intersection numbers are known. The strategy is then to calculate any $m$-form intersection number by projecting it onto $(m-1)$-form intersection numbers, and computing the remaining $1$-form intersections using a univariate algorithm.

Given an intersection number $\braket*{\varphi_{L}^{(\mathbf{m})}}{\varphi_{R}^{(\mathbf{m})}}$, we start by first projecting in terms of our $(m-1)$-form basis
\begin{equation}
    \bra*{\varphi_{L}^{(\mathbf{m})}} = \bra*{\phi_{L,i}^{(m)}}\wedge\bra*{e_{i}^{(\mathbf{m-1})}}, \quad \ket*{\varphi_{R}^{(\mathbf{m})}} = \ket*{\phi_{R,i}^{(m)}}\wedge\ket*{h_{i}^{(\mathbf{m-1})}}\,,
\end{equation}
where $\bra*{\phi_{L,i}^{(m)}}$ and $\ket*{\phi_{R,i}^{(m)}}$ are $1$-forms in the variable $z_{m}$. These projections can be calculated using the master decomposition formula (\ref{eq:masterdecomp})
\begin{equation}
    \bra*{\phi_{L,i}^{(m)}} = \sum_{j=1}^{\nu_{m-1}}\braket*{\varphi_{L}^{(\mathbf{m})}}{h_{j}^{(\mathbf{m-1})}}(\mathbf{C}^{(m-1)})^{-1}_{ji}, \quad \ket*{\phi_{R,i}^{(m)}} = \sum_{j=1}^{\nu_{m-1}}(\mathbf{C}^{(m-1)})^{-1}_{ij}\braket*{e_{j}^{(\mathbf{m-1})}}{\varphi_{R}^{(\mathbf{m})}}\,,
\end{equation}
where $(\mathbf{C}^{(m)})_{ij}=\braket*{e_{i}^{(\mathbf{m})}}{h_{j}^{(\mathbf{m})}}$ represents the $\mathbf{C}$-matrix at the $m^{th}$ layer of the iterative process. The $m$-form intersection number then reduces to
\begin{equation}
    \braket*{\varphi_{L}^{(\mathbf{m})}}{\varphi_{R}^{(\mathbf{m})}} = \sum_{p\in\mathcal{P}_{m}}\text{Res}_{z_{m}=p}\left(\psi_{L,i}(\mathbf{C}^{(m-1)})_{ij}\phi_{R,j}^{(m)}\right)\,.
\end{equation}
In direct analogy with the univariate case, $\psi_{L,i}$ is the solution to the differential equation
\begin{equation}
    \dd\psi_{L,i}+\psi_{L,j}(\mathbf{\Omega}^{(m)})_{ji} = \phi_{L,i}^{(m)}\,.
\end{equation}
In this expression, $\mathbf{\Omega}^{(m)}$ is known as the \textit{connection matrix}, and is given by the differential equation matrix for the basis $\bra*{e_{i}^{(\mathbf{m-1})}}$ in the variable $z_{m}$:
\begin{equation}
    \bra*{\grad_{\omega_{m}}e_{i}^{(\mathbf{m-1})}}:=\partial_{z_{m}}\bra*{e_{i}^{(\mathbf{m-1})}} = (\mathbf{\Omega}^{(m)})_{ij}\bra*{e_{j}^{(\mathbf{m-1})}}\,,
\end{equation}
Using the master decomposition formula (\ref{eq:masterdecomp}), $(\mathbf{\Omega}^{(m)})$ can be calculated as
\begin{align}
    (\mathbf{\Omega}^{(m)})_{ij} = \sum_{k=1}^{\nu_{m-1}}\braket*{\grad_{\omega_{m}}e_{i}^{(\mathbf{m-1})}}{h_{k}^{(\mathbf{m-1})}}(\mathbf{C}^{(m-1)})^{-1}_{kj}\,.
\end{align}
In the above expressions, in analogy to the univariate case, $\mathcal{P}_{m}$ is defined as the set of poles of the $\mathbf{\Omega}^{(m)}$ matrix. For all computations in this work, all intersection numbers were verified using this algorithm, as well as further techniques and ideas presented in \cite{brunello2023intersection}.

\subsection{Relative Twisted Cohomology}\label{subsec:relcohom}
In Baikov representation, Feynman integrals are represented as twisted integrals (see Section \ref{subsec:baikov}), but they do not satisfy the condition (\ref{eq:regulatedcondition}), namely that they are regulated integrals. This is because the differential forms have the following structure
\begin{equation}
    \varphi_L=\frac{\mathrm{d}z_1\wedge\cdots\wedge \mathrm{d}z_n}{z_1^{a_1}\cdots z_n^{a_n}}\,,
\end{equation}
but the Baikov polynomial $b$ is such that $b(\mathbf{z}=0)\neq0$, and thus $\mathbf{0}\notin\mathcal{P}({\omega})$.

An appropriate mathematical framework to deal with non regulated intersection numbers is \textit{Relative Twisted Cohomology} \cite{matsumoto2019relative,Caron_Huot_2021,Caron_Huot_2022,brunello2023intersection}. In this formalism, the contribution of the unregulated poles is captured through the usage of new dual forms $\delta$. Roughly speaking, these ``delta forms" act as residue operators, ``integrating out" variables from the intersection number. For $1$-form intersection numbers, they are defined to operate as
\begin{equation}
    \braket*{{\varphi}_L}{{\delta}_z}=\braket*{\hat{\varphi}_L \mathrm{d}z}{{\delta}_z}:= \mathrm{Res}_{z=0}\left(\frac{u(z)}{u(0)}\hat{\varphi}_L\right)\,,
\end{equation}
where we have used the notation $f=\hat{f}\,\mathrm{d}z$. As shown in \cite{brunello2023intersection}, it is also possible to consider generalisations of the delta forms involving their derivatives. For 1-forms we have
\begin{equation}\label{eq:delta_dervative}
    \braket*{\varphi_L}{\partial^{(k)}_z \delta_z}:=\frac{1}{k!}\,\Res_{z=0}\left(u(z)\,\varphi_L(z) \left(\partial_{z}^{(k)}\left.\frac{1}{u(z)}\right)\right\rvert_{z=0} \right)\,.
\end{equation}
For $n$-form intersection numbers, suppose
\begin{equation}
\begin{aligned}
    \bra{\varphi_L}&=\hat{\varphi}\,\mathrm{d}z_1\wedge\cdots\wedge \mathrm{d}z_n\,,\\
    \ket{\varphi_R}&=\hat{\phi}\,\ket{\delta_{z_1}}\wedge\dots\wedge\ket{\delta_{z_a}}\wedge \,\mathrm{d}z_{a+1}\wedge\cdots\wedge \mathrm{d}z_n\,,
\end{aligned}
\end{equation}
where $\hat{\phi}$ is any rational function in the variables $\mathbf{z}$. We then have
\begin{equation}\label{eq:relativemasterformula}
    \braket{\varphi_L}{\varphi_R}:=\braket{\mathrm{Res}_{z_1\cdots z_a=0}\left(\frac{u\,\hat{\varphi}\,dz_1\wedge dz_n}{u\rvert_{z_1\cdots z_a = 0}}\right)}{\hat{\phi}\rvert_{z_1\cdots z_a = 0}\,\mathrm{d}z_{a+1}\wedge\cdots\wedge\mathrm{d}z_n}\,.
\end{equation}
In the above expression, the multivariate residue on the variables $z_1\cdots z_a$ evaluates to a rational function. The remaining $(n-a)$-form intersection number is then computed using the recursive intersection algorithm for multivariate forms introduced above. For the rest of this work we will interchangeably use the notations $\ket{\delta_{z_1}}\wedge\dots\wedge\ket{\delta_{z_n}}=\ket{\delta_{z_1\cdots z_n}}$.

\subsection{One-loop Feynman Integrals as Twisted Integrals}\label{subsec:baikov}
Any arbitrary one-loop Feynman Integral, with $N$ external legs of momentum $p_{i}$, can be written as
\begin{equation}
    I(n_{1},\dots,n_{E+1}) = \int\frac{\dd^{d}k}{(2\pi)^{d/2}}\frac{1}{D_{1}^{n_{1}}\dots D_{N}^{n_{N}}}\,.
\end{equation}
We can view this integral as a twisted period by changing to Baikov representation \cite{Baikov:1996iu,Baikov_1997}, where it takes the form
\begin{equation}
    I \propto\int_{C_{R}}u(\mathbf{z})\,\varphi_{L}\,.
\end{equation}
The twist is given by
\begin{equation}
    u(\mathbf{z}) = b(\mathbf{z})^{\gamma}, \quad b(\mathbf{z}) = G(k,p_{1},\dots,p_{N-1})\eval_{D_{i}\rightarrow z_{i}}, \quad \gamma=\frac{d-N-1}{2}\,,
\end{equation}
where $G(\{x_{i}\})=\det[x_{i}\cdot x_{j}]$ is the Gram determinant of the loop and external momenta $\{x_i\}$, and $\varphi_L$ is a differential form given by
\begin{equation}
    \varphi_{L} = \frac{d^{N}\mathbf{z}}{z_{1}^{n_{1}}\dots z_{N}^{n_{N}}}\,.
\end{equation}
Due to the form of $G(\{x_{i}\})$, the Baikov polynomial $b(\mathbf{z})$ is always quadratic in all $\mathbf{z}$ variables. Thus, all one-loop Feynman integrals will be associated to quadratic twists. This property allows for the results presented in Sections \ref{sec:diagonal_basis_prescription} and \ref{sec:closed_formula} to be immediately applied to one-loop Feynman integrals.

\section{Diagonal Basis Prescription}\label{sec:diagonal_basis_prescription}
Our new prescription to generate diagonal $\mathbf{C}$-matrices for many one-loop Feynman Integrals is a simple extension to the algorithm presented in \cite[Section 4.4]{brunello2023intersection}. To this end we review this algorithm first, before building upon it in Section \ref{subsec:diagonal_prescription}. In Section \ref{subsec:initial examples} we showcase the algorithm on three distinct examples in detail. Beyond these, many more examples can be found in Section \ref{sec:examples}.
\subsection{Standard Basis Generation Prescription}\label{subsec:standard_prescription}
Our starting point is a twist, obtained from the Baikov polynomials $b_i(\mathbf{z})$ of a Feynman integral family, and a set of $n$ propagators $\mathbf{z}=\{z_1\cdots z_n\}$. Our goal is to find a valid basis of MIs for this integral family. In the language of relative cohomology intersection numbers, this condition translates to finding a basis such that the $\mathbf{C}$-matrix is invertible.

To this end, consider a (sub)-sector $\mathcal{S}$ of our integral family, which we represent as a set containing the propagators present in the sector
\begin{equation}
 \mathcal{S}\subseteq \{z_{1}\cdots z_{n}\} \ .
\end{equation}
To count how many master integrals are inside $\mathcal{S}$, we construct the ``regulated" twist and its respective $\omega$ form
\begin{equation}
u(\mathbf{z})=b(\mathbf{z})^\gamma\,,\qquad u_\rho(\mathbf{z})=\left(\prod_{z_i\in\mathcal{S}}z_i^{\rho_i}\right)u(\mathbf{z})\,,\qquad \omega_\rho(\mathbf{z})=\mathrm{d}\log(u_\rho(\mathbf{z})) \,.
\end{equation}
The number of solutions to the equation $\omega_\rho(\mathbf{z})=0$, denoted as $\nu_\mathcal{S}$, is the number of master integrals present in the sector $\mathcal{S}$ \cite{Lee_2013,Frellesvig_2021,brunello2023intersection}. One repeats this procedure for all sectors and subsectors to obtain a full basis with the correct number of master integrals in each sector. The end of this procedure is a basis $\tilde{e}$. We construct its dual basis counterpart by replacing inverse powers of the variables $\{z_1\cdots z_n\}$ with delta forms, namely
\begin{equation}
    \tilde{h} = \tilde{e} |_{(z_i z_j \cdots)^{-1}\rightarrow \ \delta_{z_i z_j \cdots}} \,.
\end{equation}
For an example of this algorithm in practice, see \cite[Appendix B]{brunello2023intersection}. It is worth noting that these choices of $\tilde{e}$ and $\tilde{h}$ already produce a block triangular $\mathbf{C}$-matrix. This is due to the multivariate residue operation vanishing in eq. (\ref{eq:relativemasterformula}) if there are more residues of variables than poles.

\subsection{Diagonal Basis Generation Prescription}\label{subsec:diagonal_prescription}
We now turn to the new part of this work, namely a prescription for taking $\tilde{e}$ and $\tilde{h}$ and turning them into a new basis $e$ and $h$ which produces a diagonal $\mathbf{C}$-matrix. The process is a very simple extension of the algorithm presented above, and is valid for $b(\mathbf{z})$ quadratic.

For any element in the dual basis $\tilde{h}_j$, denote by $m_j$ the number of delta forms appearing in the respective element. We then define new bases $e$, $h$ as
\begin{equation}\label{eq:diag_basis_prescription}
    e_j = \tilde{e}_j\,, \qquad h_j = \frac{\tilde{h}_j}{b(\mathbf{z})^{n-m_j}}\,.
\end{equation}
In Section \ref{subsec:1formproof}, we prove this new basis choice produces diagonal $\mathbf{C}$-matrices for $1$-form intersection numbers. Beyond $1$-forms, we conjecture that the bases $e$ and $h$ will always provide a diagonal basis for one-loop Feynman integrals with massive internal propagators and external legs respectively. In practice, we have verified this statement to be true for multiple configurations up to 6 point kinematics. For one-loop Feynman integrals with massless propagators and legs, we find that often the above prescription still works, but occasionally it fails to provide a diagonal $\mathbf{C}$-matrix. In these cases, we have observed that small changes to the bases generated by the prescription above can in many cases be implemented to once again yield diagonal $\mathbf{C}$-matrices. An example of such a situation is given in Section \ref{subsec:bhaba_single_cut} and further examples are shown in Section \ref{sec:examples}. Eq. (\ref{eq:diag_basis_prescription}) is the first new result of this work, representing a simple prescription for diagonal bases beyond $1$-form intersection numbers. 

In \cite{Caron_Huot_2021}, a similar algorithm was used to derive a basis of dual forms orthogonal to a canonical (uniform transcendental) basis of one-loop Feynman integrals. In this approach, the physical basis of Feynman integrals consists of pairs of integrals in varying space-time dimensions. We expect these results to be deeply connected, and further comparisons between these approaches are left to future works.

\subsection{First Examples}\label{subsec:initial examples}
\subsubsection{One-loop Box for Bhabha Scattering on a Double Cut}\label{subsec:bhaba_double_cut}
We illustrate the prescription of Section \ref{subsec:diagonal_prescription} for creating diagonal bases through a simple example. To this end, we return to the computation of \cite[Section 5.1]{brunello2023intersection}, namely a one-loop two mass box on the double cut.
\begin{figure}[h!]\label{fig:box2m}
\centering
\includegraphics[scale=0.4]{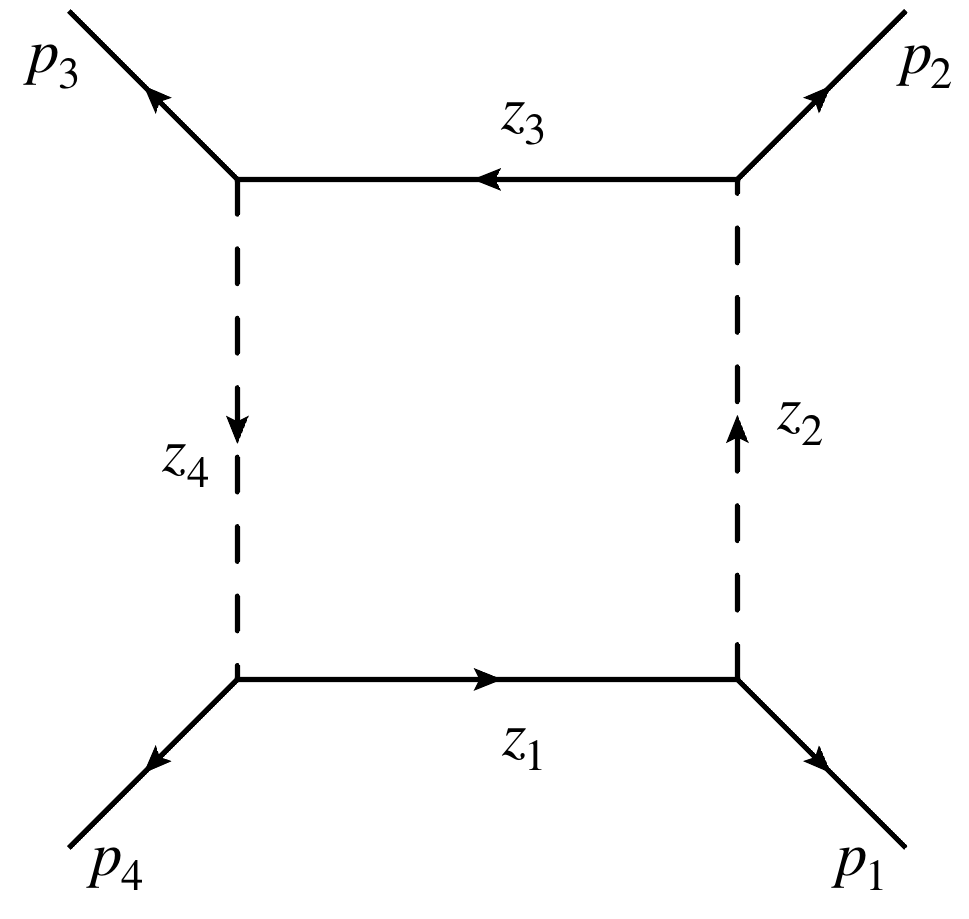}
\caption{One-loop Massive Box with two masses.}
\end{figure}

The propagators are defined as
\begin{equation}
        z_1  =  \ell^2 -m^2, \quad
        z_2  =  (\ell-p_1)^2, \quad
        z_3  =  (\ell-p_1-p_2)^2-m^2, \quad
        z_4 =  (\ell -p_1-p_2-p_3)^2,
\end{equation}
with loop momenta $\ell$, and kinematics
\begin{equation}
    p_i^2=m^2, \quad s=(p_1 + p_2)^2, \quad t=(p_2+p_3)^2, \quad s  +  t  + (p_1+p_3)^2=4 m^2 \>.
\end{equation}
Upon cutting the massless propagators $z_2$ and $z_4$, the twist reads up to a constant prefactor:
\begin{equation}
    u=b(\mathbf{z})^{(d-5)/2}, \quad b(\mathbf{z})=4 m^2 \left(s t+\left(z_1-z_3\right){}^2\right)-s^2 t+2 s t \left(z_1+z_3\right)+4 s z_1 z_3-t \left(z_1-z_3\right){}^2.
\end{equation}
Using the algorithm outlined in Section \ref{subsec:standard_prescription}, the internal and outer bases for the intersection numbers are given as:
\begin{equation}
\begin{split}
    \tilde{e}^{(3)}  = \left\{ 1 \, , \; \frac{1}{z_3} \right\}, \qquad
    \tilde{e}= \left\{ 1 \, , \; \frac{1}{z_1} \, , \; \frac{1}{z_3}  \, , \; \frac{1}{z_1 z_3} \right\}
    \>,
\end{split}
\end{equation}
with dual bases
\begin{equation}
\begin{split}
    \tilde{h}^{(3)}  = \{ 1 , \delta_3 \}, \qquad
    \tilde{h} = \{ 1 , \delta_1 , \delta_3 , \delta_{13} \}
    \>.
\end{split}
\end{equation}
To aid comparisons with the new prescription, we compute the relevant $\mathbf{C}$-matrices $\mathbf{\tilde{C}}^{(3)}=\braket*{\tilde{e}^{(3)}}{\tilde{h}^{(3)}}$ and $\mathbf{\tilde{C}}=\braket*{\tilde{e}}{\tilde{h}}$ for this basis choice. We have:
\begin{equation}\label{eq:cmat_internal_doublecutbox}
    \mathbf{\tilde{C}}^{(3)}=\left(
\begin{array}{cc}
 \frac{4 (d-5) s \left(-4 m^2+s+t\right) \left(m^2 t+z_1 \left(t+z_1\right)\right)}{(d-6) (d-4) \left(t-4 m^2\right)^2} & 0 \\
 -\frac{z_1 \left(-4 m^2+2 s+t\right)+s t}{(d-6) \left(4 m^2-t\right)} & 1 \\
\end{array}
\right) \ ,
\end{equation}
and
\begin{equation}
\mathbf{\tilde{C}}=\left(
\begin{array}{cccc}
 -\frac{s t^2 \left(-4 m^2+s+t\right)}{4 (d-7) (d-3)} & 0 & 0 & 0 \\
 \frac{s t^2 \left((44-8 d) m^2+(d-6) t\right) \left(-4 m^2+s+t\right)}{2 (d-7) (d-6) (d-4) \left(t-4 m^2\right)^2} & -\frac{4 (d-5) m^2 s t \left(4 m^2-s-t\right)}{(d-6) (d-4) \left(t-4 m^2\right)^2} & 0 & 0 \\
 \frac{s t^2 \left((44-8 d) m^2+(d-6) t\right) \left(-4 m^2+s+t\right)}{2 (d-7) (d-6) (d-4) \left(t-4 m^2\right)^2} & 0 & -\frac{4 (d-5) m^2 s t \left(4 m^2-s-t\right)}{(d-6) (d-4) \left(t-4 m^2\right)^2} & 0 \\
 \frac{s t^2}{(d-7) (d-6) \left(4 m^2-t\right)} & \frac{s t}{(d-6) \left(t-4 m^2\right)} & \frac{s t}{(d-6) \left(t-4 m^2\right)} & 1 \\
\end{array}
\right) \ .
\end{equation} \\
We now return to the bases and apply the procedure of Section \ref{subsec:diagonal_prescription} to obtain
\begin{equation}\label{eq:bhabba_doublecut_bases}
\begin{split}
    h^{(3)}  = \left\{ \frac{1}{b} , \delta_3 \right\}, \qquad
    h = \left\{ \frac{1}{b^2} , \frac{\delta_1}{b}, \frac{\delta_3}{b}, \delta_{13} \right\}
    \>\,,
\end{split}
\end{equation}
and $e^{(3)}=\tilde{e}^{(3)}\,$, $e=\tilde{e}\,$.
With this new choice the $\mathbf{C}$-matrices now read
\begin{equation}
    \mathbf{C}^{(3)}=\left(
\begin{array}{cc}
 \frac{1}{(d-4) t \left(4 m^2-t\right)} & 0 \\
 0 & 1 \\
\end{array}
\right)\,,
\end{equation}
and
\begin{equation}\label{eq:example_diag_cmat}
\mathbf{C}=\left(
\begin{array}{cccc}
 -\frac{1}{4 (d-3)^2 s t^2 \left(-4 m^2+s+t\right)} & 0 & 0 & 0 \\
 0 & \frac{1}{(d-4) t \left(4 m^2-t\right)} & 0 & 0 \\
 0 & 0 & \frac{1}{(d-4) t \left(4 m^2-t\right)} & 0 \\
 0 & 0 & 0 & 1 \\
\end{array}
\right)\,.
\end{equation} \\

Several comments are now in order. Firstly, with the new choices of $h^{(3)}$ and $h$ the $\mathbf{C}$-matrices $\mathbf{C}^{(3)}$ and $\mathbf{C}$ are now, as anticipated, diagonalised. Remarkably, $\mathbf{C}^{(3)}$ is not only diagonal, but is also independent of the external variable $z_1$, (cf eq. (\ref{eq:cmat_internal_doublecutbox})). Finally, the individual intersection numbers, both in the internal and external $\mathbf{C}$-matrices, are greatly simplified compared to the basis $\tilde{e}$ and $\tilde{h}$. These properties seem to be rather general and we have found them to hold for all one-loop examples where the inner and outer matrices have been successfully diagonalised.
\subsubsection{One-loop Massive Triangle}\label{subsec:massive_triangle}

\begin{figure}[h!]
\centering
\includegraphics[scale=0.3]{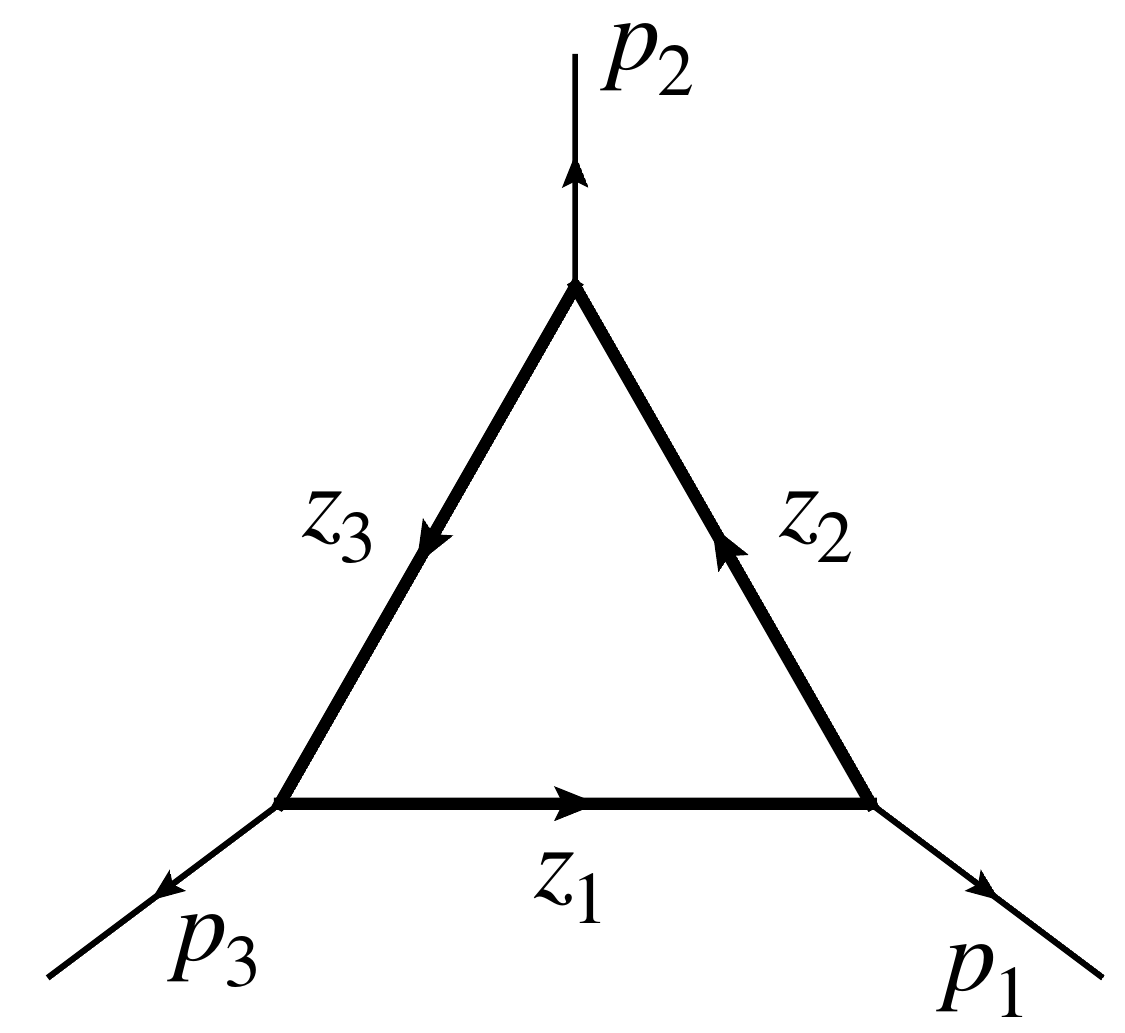}
\caption{One-loop Massive Triangle with external momenta $p_{1},p_{2},p_{3}$ and propagators $z_{1},z_{2},z_{3}$\label{fig:OLMT}.}
\end{figure}

As a second example, let us consider a one-loop triangle with massive propagators and external legs shown in Figure \ref{fig:OLMT}. 

We define the propagators as
\begin{equation}
    z_1=l^2-m_1^2\,, \qquad z_2 = \left(l+p_1\right){}^2-m_1^2\,, \qquad z_3 = \left(l+p_1+p_2\right){}^2-m_1^2\,,
\end{equation}
with kinematics
\begin{equation}
    p_i^2 = m_2^2\,, \qquad (p_1+p_2)^2=m_2^2 \,.
\end{equation}
The twist reads
\begin{equation}
    u = b(\mathbf{z})^{\frac{d-4}{2}}\,, \quad b(\mathbf{z})=m_2^2 \left(z_1+z_2+z_3\right)-m_2^4+3 m_1^2 m_2^2-z_1^2-z_2^2-z_3^2+z_1 z_2+\left(z_1+z_2\right) z_3\,.
\end{equation}
We pick the variable ordering $\left\{z_3\,,z_2\,,z_1\right\}$ and, using the algorithm described in Section \ref{subsec:standard_prescription} pick bases
\begin{equation}
\begin{aligned}
    \tilde{e}^{(1)}&=\left\{1,\frac{1}{z_1}\right\} \,,\qquad \tilde{e}^{(21)}= \left\{1,\frac{1}{z_1},\frac{1}{z_2},\frac{1}{z_1 z_2}\right\}\,,\qquad \tilde{e}= \left\{\frac{1}{z_1},\frac{1}{z_2},\frac{1}{z_3},\frac{1}{z_1 z_2},\frac{1}{z_1 z_3},\frac{1}{z_2 z_3},\frac{1}{z_1 z_2 z_3}\right\}\,,\qquad \\
    \tilde{h}^{(1)}&=\left\{1,\delta_1\right\}\,, \qquad \,\,\,\, \tilde{h}^{(21)}=\left\{1,\delta_1,\delta_2,\delta_{12}\right\}\,, \qquad \quad \,\,\,\,\tilde{h}=\left\{\delta_1,\delta_2,\delta_3,\delta_{12},\delta_{13},\delta_{23},\delta_{123}\right\}\,.
\end{aligned}
\end{equation}
Following the diagonal basis prescription we generate the new basis
\begin{equation}
\begin{aligned}
    {e}^{(1)}&=\left\{1,\frac{1}{z_1}\right\} \,,\qquad {e}^{(21)}= \left\{1,\frac{1}{z_1},\frac{1}{z_2},\frac{1}{z_1 z_2}\right\}\,,\qquad {e}= \left\{\frac{1}{z_1},\frac{1}{z_2},\frac{1}{z_3},\frac{1}{z_1 z_2},\frac{1}{z_1 z_3},\frac{1}{z_2 z_3},\frac{1}{z_1 z_2 z_3}\right\}\,,\qquad \\
    {h}^{(1)}&=\left\{\frac{1}{b},\delta_1\right\}\,, \quad\,\,\,\,\, {h}^{(21)}=\left\{\frac{1}{b^2},\frac{\delta_1}{b},\frac{\delta_2}{b},\delta_{12}\right\}\,, \quad\,\,\,\,\,\,\,{h}=\left\{\frac{\delta_1}{b^2},\frac{\delta_2}{b^2},\frac{\delta_3}{b^2},\frac{\delta_{12}}{b},\frac{\delta_{13}}{b},\frac{\delta_{23}}{b},\delta_{123}\right\}\,.
\end{aligned}
\end{equation}
With this basis, the $\mathbf{C}$-matrices read
\begin{equation}
\begin{aligned}
    \mathbf{C}^{(1)}&=\left(
\begin{array}{cc}
 \frac{1}{(d-3) m_2^2} & 0 \\
 0 & 1 \\
\end{array}
\right)\,,\\
    \mathbf{C}^{(21)}&=\left(
\begin{array}{cccc}
 \frac{4}{3 (d-2)^2 m_2^4} & 0 & 0 & 0 \\
 0 & \frac{1}{(d-3) m_2^2} & 0 & 0 \\
 0 & 0 & \frac{1}{(d-3) m_2^2} & 0 \\
 0 & 0 & 0 & 1 \\
\end{array}
\right)\,,\\
    \mathbf{C}&=\left(
\begin{array}{ccccccc}
 \frac{4}{3 (d-2)^2 m_2^4} & 0 & 0 & 0 & 0 & 0 & 0 \\
 0 & \frac{4}{3 (d-2)^2 m_2^4} & 0 & 0 & 0 & 0 & 0 \\
 0 & 0 & \frac{4}{3 (d-2)^2 m_2^4} & 0 & 0 & 0 & 0 \\
 0 & 0 & 0 & \frac{1}{(d-3) m_2^2} & 0 & 0 & 0 \\
 0 & 0 & 0 & 0 & \frac{1}{(d-3) m_2^2} & 0 & 0 \\
 0 & 0 & 0 & 0 & 0 & \frac{1}{(d-3) m_2^2} & 0 \\
 0 & 0 & 0 & 0 & 0 & 0 & 1 \\
\end{array}
\right)\,.
\end{aligned}
\end{equation}
Once again, each $\mathbf{C}$-matrix for each layer is diagonalised, and the internal $\mathbf{C}$-matrices do not depend on the outer variables. The intersection numbers themselves also evaluate to very simple functions.
\subsubsection{One-loop Box for Bhabha Scattering on a Single Cut}\label{subsec:bhaba_single_cut}
To illustrate a case where massless propagators require the diagonal basis prescription algorithm to be slightly modified, we return to the example of Section \ref{subsec:bhaba_double_cut}, but this time consider a different cut, $z_1=0$. By choosing this cut, we have two massless propagators, $z_2$ and $z_4$. The twist reads
\begin{equation}
\begin{split}
    u &= b(\mathbf{z})^{\frac{d-5}{2}}\,, \\
    b(\mathbf{z}) &= s^2 z_4^2+\left(t \left(z_3-s\right)+s z_2\right){}^2-2 s z_4 \left(z_2 (s+2 t)+t \left(s-z_3\right)\right)\\
    &\quad -4 m^2 \left(s \left(-2 z_2 \left(t+z_4\right)+\left(t-z_4\right){}^2+z_2^2\right)+t z_3^2\right)\,.
\end{split}
\end{equation}
We pick the variable ordering $\left\{z_4\,,z_3\,,z_2\right\}$ and using the algorithm in Section \ref{subsec:standard_prescription} pick bases
\begin{equation}
\begin{aligned}
    \tilde{e}^{(2)}&=\left\{1,\frac{1}{z_2}\right\}\,,\qquad \tilde{e}^{(32)}=\left\{1,\frac{1}{z_2},\frac{1}{z_3},\frac{1}{z_2 z_3}\right\}\,,\qquad \tilde{e}=\left\{1,\frac{1}{z_3},\frac{1}{z_2 z_4},\frac{1}{z_2 z_3 z_4}\right\}\,,\\
    \tilde{h}^{(2)}&=\left\{1,\delta_2\right\}\,,\qquad\,\,\,\, \tilde{h}^{(32)}=\left\{1,\delta_2,\delta_3,\delta_{23}\right\}\,,\qquad \quad\,\,\,\, \tilde{h}=\left\{1,\delta_3,\delta_{24},\delta_{234}\right\}\,.
\end{aligned}
\end{equation}
Using the diagonal basis prescription, we would pick the bases $e$ and $h$ as
\begin{equation}
\begin{aligned}
    {e}^{(2)}&=\left\{1,\frac{1}{z_2}\right\}\,,\qquad {e}^{(32)}=\left\{1,\frac{1}{z_2},\frac{1}{z_3},\frac{1}{z_2 z_3}\right\}\,,\qquad {e}=\left\{1,\frac{1}{z_3},\frac{1}{z_2 z_4},\frac{1}{z_2 z_3 z_4}\right\}\,,\\
    {h}^{(2)}&=\left\{\frac{1}{b},\delta_2\right\}\,,\quad\,\,\, {h}^{(32)}=\left\{\frac{1}{b^2},\frac{\delta_2}{b},\frac{\delta_3}{b},\delta_{23}\right\}\,,\qquad {h}=\left\{\frac{1}{b^3},\frac{\delta_3}{b^2},\frac{\delta_{24}}{b},\delta_{234}\right\}\,.
\end{aligned}
\end{equation}
However, this produces an \textit{almost} diagonal basis. Explicitly, we have
\begin{equation}
\begin{aligned}
    \mathbf{C}^{(2)}&=\left(
\begin{array}{cc}
 \frac{1}{(d-4) s \left(4 m^2-s\right)} & 0 \\
 0 & 1 \\
\end{array}
\right)\,,\\
    \mathbf{C}^{(32)}&=\left(
\begin{array}{cccc}
 -\frac{1}{4 (d-3)^2 m^2 s t \left(-4 m^2+s+t\right)} & 0 & 0 & 0 \\
 0 & \frac{1}{(d-4) t \left(4 m^2-t\right)} & 0 & 0 \\
 0 & 0 & \frac{1}{(d-4) s \left(4 m^2-s\right)} & 0 \\
 0 & 0 & 0 & 1 \\
\end{array}
\right)\,,\\
    \mathbf{C}&=\left(
\begin{array}{cccc}
 \frac{1}{4 (d-3)(d-2)(d-1) s^2 t^2 \left(-4 m^2+s+t\right)^2} & 0 & 0 & 0 \\
 0 & -\frac{s-4 m^2}{4 (d-5)(d-3) s} & 0 & 0 \\
 \frac{1}{4 (d-3)^2 (d-1) m^2 s^2 t^3 \left(-4 m^2+s+t\right)^2} & 0 & \frac{1}{(d-4) t \left(4 m^2-t\right)} & 0 \\
 \frac{7-2 d}{4 (d-4) (d-3)^2 (d-1) m^2 s^2 t^3 \left(4 m^2-s\right) \left(-4 m^2+s+t\right)^2} & 0 & 0 & 1 \\
\end{array}
\right)\,.
\end{aligned}
\end{equation}
To produce a diagonal basis, we modify $e$ and $h$ to
\begin{equation}\label{eq:massless_modification}
{e}=\left\{1,\frac{1}{z_3}\frac{1}{b},\frac{1}{z_2 z_4},\frac{1}{z_2 z_3 z_4}\right\}\,,\qquad {h}=\left\{\frac{1}{b^2},\frac{\delta_3}{b},\frac{\delta_{24}}{b},\delta_{234}\right\} \,,
\end{equation}
keeping the internal bases unchanged. We note that this choice, unlike the previous examples, requires a modification of the left basis $e$. With this new choice, we once again obtain a diagonal outer matrix
\begin{equation}
    \mathbf{C}=\left(
\begin{array}{cccc}
 -\frac{m^2}{(d-4) (d-3) (d-2) s t \left(-4 m^2+s+t\right)} & 0 & 0 & 0 \\
 0 & -\frac{1}{4 (d-5)^2 s^2 t \left(-4 m^2+s+t\right)} & 0 & 0 \\
 0 & 0 & \frac{1}{(d-4) t \left(4 m^2-t\right)} & 0 \\
 0 & 0 & 0 & 1 \\
\end{array}
\right)\,.
\end{equation}
In general, we find that some one-loop diagrams with massless propagators require modifications of this type to the diagonal basis algorithm to obtain a fully diagonal outer basis. 

Section \ref{sec:examples} contains many more examples of the diagonal basis algorithm in use.

\subsection{Proof for 1-form Intersection Numbers}\label{subsec:1formproof}

An algorithm for producing a diagonal basis for any univariate twist was presented in \cite{Frellesvig_2019}. Here we present a different approach specifically suited to the prescription of Section \ref{subsec:diagonal_prescription}: As outlined in Section \ref{subsec:baikov}, all one-loop Feynman integrals produce quadratic Baikov polynomials. To this end, we consider a generic quadratic univariate polynomial and it's respective twist
\begin{equation}
    b(z) = a_0 + a_1 z + a_2 z^2 = a_2\,(z-r_1)(z-r_2)\,,\quad u(z)=b(z)^\gamma\,,\quad \omega(z)=\gamma  \left(\frac{1}{z-r_1}+\frac{1}{z-r_2}\right)\mathrm{d}z\,,
\end{equation}
where $r_1\neq r_2\neq 0$ and $\gamma$ is an generic constant. As in the Baikov representation, we interpret $z$ to be a propagator, and thus are interested in constructing a basis for the sector $\mathcal{S}_{1}=\{z\}$ and its subsector $\mathcal{S}_{0}=\emptyset$. Using the algorithm of Section \ref{subsec:standard_prescription}, we find $\mathcal{S}_{0}$ to have one master integral\footnote{For twists originating from physical Feynman integrals, this unphysical subsector, corresponding to a Feynman integral without propagators, contains no master integrals.} and $\mathcal{S}=\{z\}$ to have two. Thus, we pick bases
\begin{equation}
    \tilde{e}=\left\{1\,,\frac{1}{z}\right\}\,,\qquad \tilde{h}=\left\{1\,,\delta_{z}\right\}\,,
\end{equation}
and consequently
\begin{equation}
    {e}=\left\{1\,,\frac{1}{z}\right\}\,,\qquad {h}=\left\{\frac{1}{b(z)}\,,\delta_{z}\right\}\,.
\end{equation}
The relevant $\mathbf{C}$-matrix is
\begin{equation}
    \mathbf{C}_{ij}=\braket{e_i\,\dd z}{h_j\,\dd z}=\left(
\begin{array}{cc}
 c_{11} & c_{12} \\
 c_{21} & c_{22} \\
\end{array}
\right)
\end{equation}
From eq.\,\,(\ref{eq:relativemasterformula}) we have $c_{22}=1$ and $c_{12}=0$. Thus, it remains to show that $c_{21}=\braket{e_2}{h_1}=\braket{1/z}{1/b(z)}=0$. To do this we analyse the poles of the two forms appearing. We have
\begin{equation}
    \mathcal{P}(e_2)=\{0\,,\infty\}\,,\qquad \mathcal{P}(h_1)=\{r_1,r_2\}\,.
\end{equation}
Furthermore, we note that all poles are simple, and as such $e_2$ and $h_1$ can be written as $\mathrm{d}\log$ forms. We can thus leverage a known formula for $\mathrm{d}\log$ intersection numbers in hyperplanar arrangements\footnote{For univariate intersection numbers, the arrangement is always hyperplanar.}\,\cite{ojm/1200788347}.
\begin{equation}
    \braket{\varphi_L}{\varphi_R}\propto \sum_{\{r_i,r_j\}\, \in\, \mathcal{P}(\varphi_L)\cap \mathcal{P}(\varphi_R)}\frac{1}{\gamma_i\,\gamma_j}\,.
\end{equation}
Instead of explaining this result in detail, we jump straight to the point. For $\braket{e_2}{h_1}$ we have
\begin{equation}
    \mathcal{P}(e_2)\cap\mathcal{P}(h_1)=\emptyset\,,
\end{equation}
and thus there are no terms in the sum, implying $c_{21}=0$. For more details, see \cite{ojm/1200788347,mizerascatt}.

It is worth nothing that this argument does not imply that $c_{11}=0$, because $1\,\dd z$ is not $\dd\log$ due to its double pole at $\infty$. Additionally, this argument still shows $\braket{1/z}{1/b(z)}=0$ for higher order polynomials $b(z)$ beyond quadratics. However, in these cases, the sector $\mathcal{S}_{1}$ will include more master integrals, and thus just $c_{21}$ vanishing would not guarantee a diagonal basis.

\section{Closed Formula for Intersection Numbers \label{sec:closed_formula}}

In Section \ref{sec:diagonal_basis_prescription} a prescription was introduced for diagonalising intersection numbers related to quadratic twists $u(\mathbf{z})=b(\mathbf{z})^\gamma$. In this section, we focus on the properties of intersection numbers generated by this prescription, specifically of the form $\braket{b(\mathbf{z})^p}{b(\mathbf{z})^q}$\footnote{For the case $p=q=1$ a similar result was derived in \cite[Section 4 \& Appendix B]{Caron_Huot_2021}}. We find that such intersection numbers follow very regular patterns, and in Section \ref{subsec:closed_formula} propose a closed form formula for their evaluation, for any $p\,,\,q$ integer. To the author's knowledge, this is the first such proposition beyond $\mathrm{d}\log$ forms and as such constitutes the second main result of this work.

In Section \ref{subsec:diagonal_terms_of_cmatrix} we show that all the intersection numbers appearing in the diagonal basis prescription are of this form. Thus, the full $\mathbf{C}$-matrix can be computed with just this closed formula, without the need of any algorithmic evaluation strategies.

\subsection{Closed Formula For Intersection Numbers}\label{subsec:closed_formula}
We consider intersection numbers of the kind
\begin{equation}
    u(\mathbf{z})=b(\mathbf{z})^\gamma\,,\qquad\braket{b(\mathbf{z})^p}{b(\mathbf{z})^q} \,,
\end{equation}
where we stress again $b(\mathbf{z})$ is a quadratic polynomial in $n$ variables. The closed formula for this intersection number reads:
\begin{equation}\label{eq:closed_formula}
    \braket{b(\mathbf{z})^p}{b(\mathbf{z})^q}=f_{n}(p,q;\gamma)\times\frac{\det\left(\mathbf{H}(b_h)\right)^{n+p+q}}{\det\left(\mathbf{H}(b)\right)^{n+p+q+1}}\,,
\end{equation}
In this expression, $b_h$ is the polynomial obtained by homogenising $b$ with an extra variable $z_{n+1}$, and $\mathbf{H}$ is the Hessian matrix computed with respect to the variables $\mathbf{z}$ for $b$ and $\{\mathbf{z},z_{n+1}\}$ for $b_h$. Since $b$ and $b_{h}$ are quadratic, their Hessian matrices do not depend on the variables $\mathbf{z}$. In Appendix \ref{app:multivariate_discriminant}, we show that $\det\left(\mathbf{H}(b_h)\right)$, up to a numerical prefactor, is equivalent to the \textit{multivariate discriminant} of $b$. Finally, $f_{n}(p,q;\gamma)$ is a $\gamma$ dependent prefactor given by
\begin{equation}
    f_{n}(p,q;\gamma)=\frac{(-1)^n}{(\gamma+p+n/2)}\left(\prod_{i=1}^{n-1}\frac{1}{(\gamma+p+i)}\right)\left(\prod _{i=1}^{n+p+q} \frac{\gamma+p -i+n}{2 \gamma+2p -2 i+n}\right)\left(\prod _{i=0}^{-n-p-q-1} \frac{2 \gamma+2p +2 i+n}{\gamma+p +i+n}\right)\,.
\end{equation}
\subsection{Application to the $\mathbf{C}$-matrix}\label{subsec:diagonal_terms_of_cmatrix}

The formula presented above can be immediately put to use in computing $\mathbf{C}$-matrices. In general, using the construction of Sections \ref{subsec:standard_prescription} and \ref{subsec:diagonal_prescription}, the on diagonal elements of the $\mathbf{C}$-matrix are $n$-form intersection numbers of the type
\begin{equation}\label{eq:closed_formula_cmat}
    \mathbf{C}_{ii}=\braket{e_i}{h_i}=\braket{\left(\prod_{z_j\in \mathcal{S}_i}\frac{1}{z_j}\right)b(\mathbf{z})^{p}}{\left(\prod_{z_j\in \mathcal{S}_i}\delta_j\right)b(\mathbf{z})^q}\,,
\end{equation}
where each $\mathcal{S}_i$ is a (sub)sector of the integral corresponding to the variables appearing in the denominator of $e_{i}$. The quantities $p$ and $q$ are kept as generic integer coefficients, to allow for cases such as eq. (\ref{eq:massless_modification}). From eq. (\ref{eq:relativemasterformula}) the delta functions are immediately integrated out: If the sector $\mathcal{S}_i$ has $k$ propagators, the dual basis will contain $k$ delta forms, and thus upon integrating them out we obtain an $(n-k)$-form intersection number of the form
\begin{equation}
    \mathbf{C}_{ii}=\braket{b(\mathbf{z})^{p}\rvert_{z_j\in \mathcal{S}_i\to 0}}{b(\mathbf{z})^{q}\rvert_{z_j\in \mathcal{S}_i\to 0}}\,.
\end{equation}
The polynomial $b(\mathbf{z})\rvert_{z_j\in \mathcal{S}_i\to 0}$ is the polynomial $b(\mathbf{z})$ with all $k$ variables appearing in the sector $\mathcal{S}_i$ set to zero. After this operation one is left still with a quadratic polynomial but in fewer variables. Thus eq. (\ref{eq:closed_formula}) can be used immediately for their computation.

By combining these results with the prescription presented in Section \ref{sec:diagonal_basis_prescription}, one can form an orthonormal basis, similarly to \cite{Caron_Huot_2021}, by normalising the dual forms $h_{j}$ by the result of eq. (\ref{eq:closed_formula}). This procedure results in $\mathbf{C}=\mathbb{I}$ and eq. (\ref{eq:masterdecomp}) simplifies to
\begin{equation}
    c_i=\braket{\varphi_L}{h_i}\,.
\end{equation}

\subsection{Example Applications}
To illustrate eq. (\ref{eq:closed_formula}), let us recompute the intersection number
\begin{equation}
    \braket{e_1}{h_1}=\braket{1}{\frac{1}{b^2}}\,.
\end{equation}
From eq. (\ref{eq:bhabba_doublecut_bases}). We have $n=2$, $p=0$, $q=-2$ and $\gamma=(d-5)/2$. The various pieces in the formula evaluate to
\begin{equation}
\begin{aligned}
    f_{2}\left(0,-2;\frac{d-5}{2}\right)&=\frac{4}{(d-3)^2}\,,\\
    \det\left(\mathbf{H}(b)\right)&=-16 s t^2 (-4 m^2 + s + t)\,,\\
    \det\left(\mathbf{H}(b_h)\right)&=-32 s^2 t^4 (-4 m^2 + s + t)^2\,,\\
\end{aligned}
\end{equation}
and thus
\begin{equation}
    \braket*{e_1}{h_1}=f_{2}\left(0,-2;\frac{d-5}{2}\right)\times\frac{1}{\det\left(\mathbf{H}(b)\right)}=-\frac{1}{4 (d-3)^2 s t^2 \left(-4 m^2+s+t\right)}\,,
\end{equation}
in agreement with eq. (\ref{eq:example_diag_cmat}). It is worth noting that for $n$-form intersection numbers of the form $\braket{1}{1/b^n}$ the more complicated polynomial $\det\left(\mathbf{H}(b_{h})\right)$ never appears in the final result, as it is always raised to the 0\textsuperscript{th} power. In cases where the diagonal basis prescription requires no modification, the intersection numbers will always be of this form.

To showcase the true power of this result, the evaluation of any other intersection numbers with large powers of $p$ and $q$ require very little extra computational effort. For example, consider $\braket*{b^{10}}{b^{-16}}$, which immediately evaluates to
\begin{equation}
    \braket*{b^{10}}{b^{-16}}=f_{2}\left(10,-16;\frac{d-5}{2}\right)\times \frac{\det\left(\mathbf{H}(b_h)\right)^{-4}}{\det\left(\mathbf{H}(b)\right)^{-3}}=\frac{64}{(d+17) (d+25)}\times \frac{-1}{256 s^5 t^{10} \left(-4 m^2+s+t\right)^5}\,,
\end{equation}
which we independently verify to be the correct result.

\section{Examples}\label{sec:examples}

In Sections \ref{sec:diagonal_basis_prescription} and \ref{sec:closed_formula} we introduced a prescription for generating an orthogonal basis of differential forms related to one-loop Feynman Integrals, and we proposed a closed form formula for the direct computation of the diagonal $\mathbf{C}$-matrix given by this basis. In this section we will present a series of one-loop examples for which we have found an orthogonal basis, and have verified that the closed formula gives all intersection numbers in the $\mathbf{C}$-matrices correctly. For the fully massive diagrams the diagonal basis comes from the prescription described in Section \ref{sec:diagonal_basis_prescription}. As anticipated previously, for the massless diagrams, we find that we must modify this prescription slightly in order to obtain an orthogonal basis. When drawing diagrams in this section, dotted lines represent massless propagators, thin lines represent masses $m_{1}$, and thick lines represent masses $m_{2}$. For the simpler examples we show explicitly the Baikov polynomial, for the more complicated examples we omit the Baikov polynomial due to its length, and consider a spanning set of cuts to simplify the computation. Finally, we always consider the external momenta to be $d$-dimensional.

\subsection{Massive Bubble}

\begin{figure}[h!]
\centering
\includegraphics[scale=0.3]{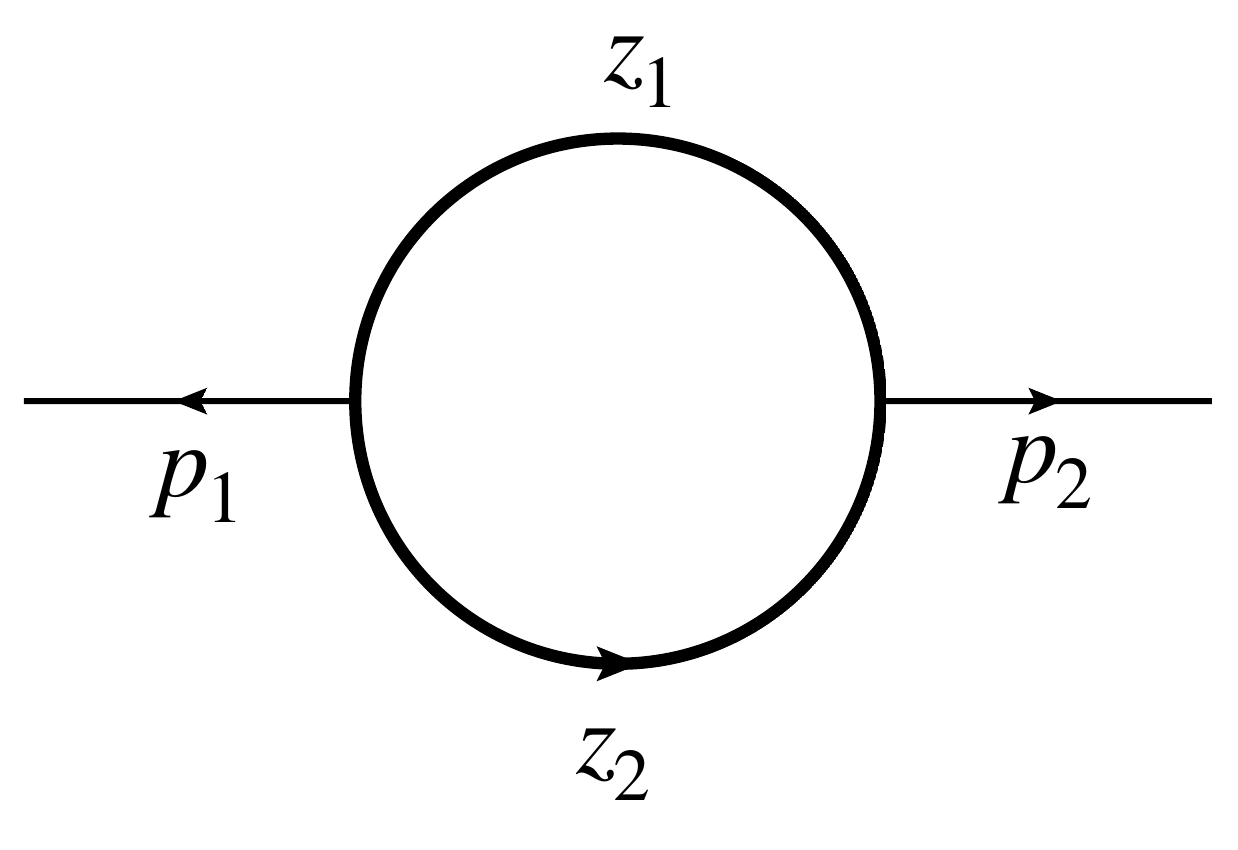}
\caption{One-loop Massive Bubble with external momenta $p_{1},p_{2}$ and propagators $z_{1},z_{2}$.\label{fig:OLMBu}}
\end{figure}

The one-loop massive bubble is shown in Fig. \ref{fig:OLMBu}

The propagators are
\begin{align}
    z_{1} = \ell^{2}-m_{2}^{2}\,, \quad z_{2} &= (\ell-p_{1})^{2}-m_{2}^{2}\,.
\end{align}
The kinematics for this problem are $p_{i}^{2}=m_{1}^{2}$ and the Baikov polynomial is
\begin{align}
    b(\mathbf{z}) = 4 m_{2}^2 m_{1}^{2}-m_{1}^4+2 m_{1}^{2} (z_{1}+z_{2})-(z_{1}-z_{2})^2\,.
\end{align}
The diagonal basis is given by
\begin{equation}
\begin{aligned}
    e = \left\{\frac{1}{z_{1}},\frac{1}{z_{2}},\frac{1}{z_{1}z_{2}}\right\}\,, \quad h = \left\{\frac{\delta_1}{b},\frac{\delta_2}{b},\delta_{12},\right\}\,.
\end{aligned}
\end{equation}

\newpage

\subsection{Massive Triangle}

The massive triangle is discussed in detail in Section \ref{subsec:massive_triangle}.

\subsection{Massless Box}

\begin{figure}[h!]
\centering
\includegraphics[scale=0.3]{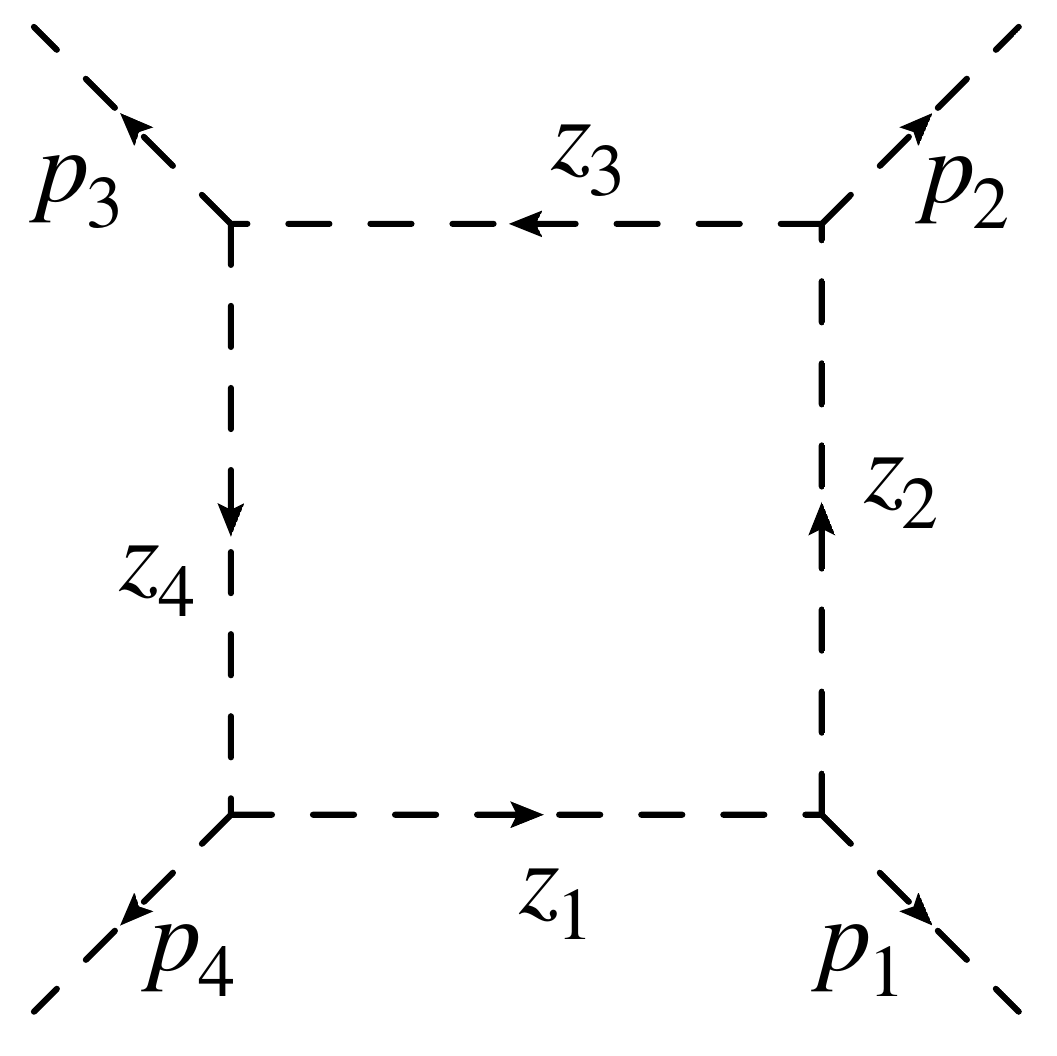}
\caption{One-loop Massless Box with external momenta $p_{1},p_{2},p_{3},p_{4}$ and propagators $z_{1},z_{2},z_{3},z_{4}$.\label{fig:OLMaBo}}
\end{figure}

The massless box is shown in Fig. \ref{fig:OLMaBo}

The propagators are
\begin{equation}
\begin{aligned}
    z_{1} = \ell^{2}, \quad z_{2} &= (\ell-p_{1})^{2}, \quad z_{3} = (\ell-p_{1}-p_{2})^{2}\,,\\
    z_{4} &= (\ell-p_{1}-p_{2}-p_{3})^{2}\,.
\end{aligned}
\end{equation}
The kinematics are
\begin{align}
    p_{i}^{2} = 0\,, \quad s = (p_{1}+p_{2})^{2}\,, \quad t = (p_{2}+p_{3})^{2}\,, \quad s+t+(p_{1}+p_{3})^{2} = 0\,.
\end{align}
The Baikov polynomial is
\begin{equation}
\begin{aligned}
    b(\mathbf{z}) &= s^2 \left(t^2-2 t (z_{2}+z_{4})+(z_{2}-z_{4})^2\right)+2 s t (-t (z_{1}+z_{3})\\
    &\quad+z_{4} (z_{1}-2 z_{2}+z_{3})+z_{1} z_{2}-2 z_{1} z_{3}+z_{2} z_{3})+t^2 (z_{1}-z_{3})^2\,.
\end{aligned}
\end{equation}
The diagonal basis is given by
\begin{equation}
\begin{aligned}
    e = \left\{\frac{1}{z_{1}z_{3}},\frac{1}{z_{2}z_{4}},\frac{1}{z_{1}z_{2}z_{3}z_{4}}\right\}\,, \quad h = \left\{\frac{\delta_{13}}{b},\frac{\delta_{24}}{b},\delta_{1234}\right\}\,.
    \end{aligned}
\end{equation}

\subsection{Massive Box}

\begin{figure}[h!]
\centering
\includegraphics[scale=0.3]{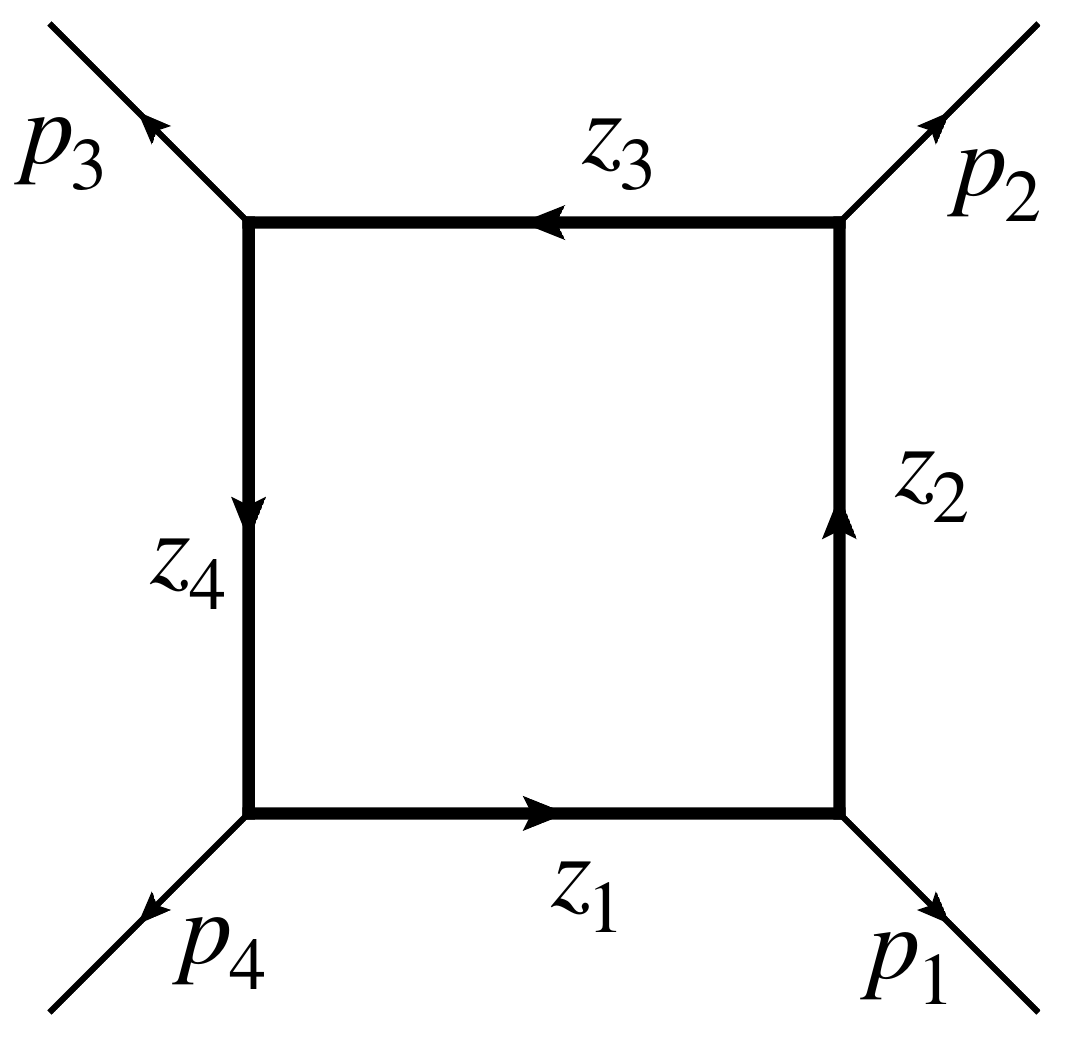}
\caption{One-loop Massive Box with external momenta $p_{1},p_{2},p_{3},p_{4}$ and propagators $z_{1},z_{2},z_{3},z_{4}$.\label{fig:OLMBo}}
\end{figure}

The massive box is shown in Fig. \ref{fig:OLMBo}

The propagators are
\begin{equation}
\begin{aligned}
    z_{1} = \ell^{2}-m_{2}^{2}, \quad z_{2} &= (\ell-p_{1})^{2}-m_{2}^{2}, \quad z_{3} = (\ell-p_{1}-p_{2})^{2}-m_{2}^{2}\,,\\
    z_{4} &= (\ell-p_{1}-p_{2}-p_{3})^{2}-m_{2}^{2}\,.
\end{aligned}
\end{equation}
The kinematics are
\begin{align}
    p_{i}^{2} = m_{1}^{2}\,, \quad s = (p_{1}+p_{2})^{2}\,, \quad t = (p_{2}+p_{3})^{2}\,, \quad s+t+(p_{1}+p_{3})^{2} = 4m_{1}^{2}\,.
\end{align}
The Baikov polynomial is
\begin{equation}
    \begin{aligned}
        b(\mathbf{z}) &= -4 m_{1}^2 s t \left(-4 m_{2}^2+s+t\right)-4 m_{2}^4 s t-4 s t z_{1} z_{3}+2 s t z_{2} z_{3}+t^2 z_{1}^2-2 t^2 z_{1} z_{3}+t^2 z_{3}^2\\
        &\quad+4 m_{2}^2 \left(s t (z_{1}+z_{2}+z_{3}+z_{4})-s (z_{2}-z_{4})^2-t (z_{1}-z_{3})^2\right)+s^2 t^2-2 s^2 t z_{2}\\
        &\quad+s^2 z_{2}^2+s^2 z_{4}^2-2 s t^2 z_{1}-2 s t^2 z_{3}-2 s z_{4} (s (t+z_{2})-t (z_{1}-2 z_{2}+z_{3}))+2 s t z_{1} z_{2}\,.
    \end{aligned}
\end{equation}
The diagonal basis is given by
\begin{equation}
\begin{aligned}
    e &= \bigg\{\frac{1}{z_{1}},\frac{1}{z_{2}},\frac{1}{z_{3}},\frac{1}{z_{4}},\frac{1}{z_{1} z_{2}},\frac{1}{z_{1} z_{3}},\frac{1}{z_{1} z_{4}},\frac{1}{z_{2} z_{3}},\frac{1}{z_{2} z_{4}},\frac{1}{z_{3} z_{4}},\\
    &\qquad\frac{1}{z_{1} z_{2} z_{3}},\frac{1}{z_{1} z_{2} z_{4}},\frac{1}{z_{1} z_{3} z_{4}},\frac{1}{z_{2} z_{3} z_{4}},\frac{1}{z_{1} z_{2} z_{3} z_{4}}\bigg\}\,,\\
    h &= \bigg\{\frac{\delta_{1}}{b^3},\frac{\delta_{2}}{b^3},\frac{\delta_{3}}{b^3},\frac{\delta_{4}}{b^3},\frac{\delta_{12}}{b^2},\frac{\delta_{13}}{b^2},\frac{\delta_{14}}{b^2},\frac{\delta_{23}}{b^2},\frac{\delta_{24}}{b^2},\frac{\delta_{34}}{b^2},\frac{\delta_{123}}{b},\frac{\delta_{124}}{b},\frac{\delta_{134}}{b},\frac{\delta_{234}}{b},\delta_{1234}\bigg\}\,.
\end{aligned}
\end{equation}

\newpage

\subsection{Massless Pentagon}

\begin{figure}[h!]
\centering
\includegraphics[scale=0.3]{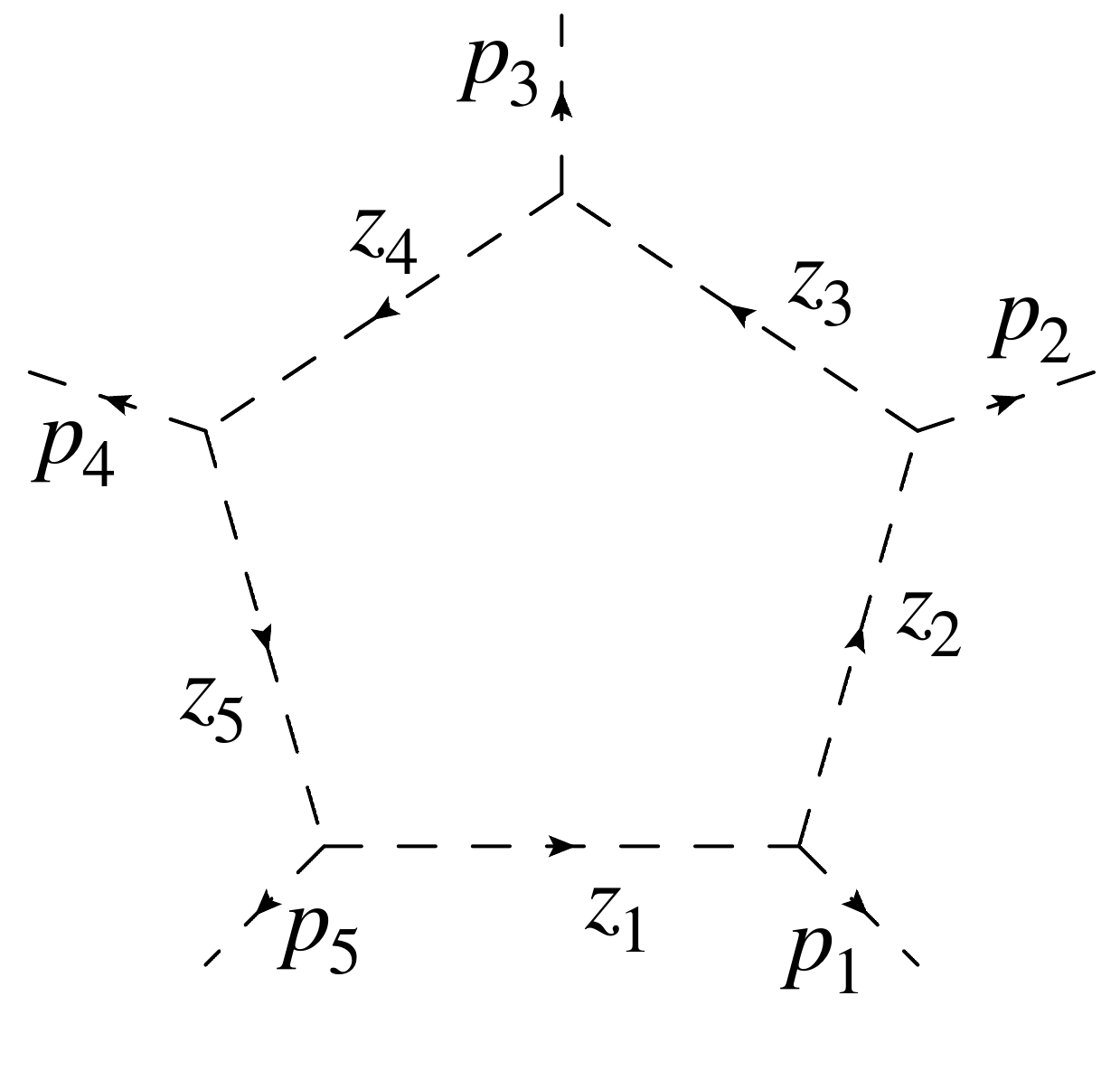}
\caption{One-loop Massless Pentagon with external momenta $p_{1},p_{2},p_{3},p_{4},p_{5}$ and propagators $z_{1},z_{2},z_{3},z_{4},z_{5}$.\label{fig:OLMaP}}
\end{figure}

The massless pentagon is shown in Fig. \ref{fig:OLMaP}

The propagators are
\begin{equation}
\begin{aligned}
    z_{1} &= \ell^{2}\,, \quad z_{2} = (\ell-p_{1})^{2}\,, \quad z_{3} = (\ell-p_{1}-p_{2})^{2}\,,\\
    z_{4} &= (\ell-p_{1}-p_{2}-p_{3})^{2}\,, \quad z_{5} = (\ell-p_{1}-p_{2}-p_{3}-p_{4})^{2}\,.
\end{aligned}
\end{equation}
The kinematics are
\begin{align}
    p_{i}^{2} = 0\,, \quad s_{ij} = (p_{i}+p_{j})^{2}\,, \quad \sum_{0<i<j<5}s_{ij} = 0\,.
\end{align}
The spanning set of cuts up to symmetry relations is
\begin{align}
    \{z_{4},z_{5}\}\,, \quad \{z_{3},z_{5}\}\,.
\end{align}

\subsubsection{Cut $\{z_{4},z_{5}\}$}

On this cut, the diagonal basis is given by
\begin{equation}
\begin{aligned}
    e = \left\{\frac{1}{z_{1} z_{2}},\frac{1}{z_{1} z_{3}},\frac{1}{z_{2} z_{3}},\frac{1}{z_{1} z_{2} z_{3}}\right\}\,, \quad h = \left\{\frac{\delta_{12}}{b},\frac{\delta_{13}}{b},\frac{\delta_{23}}{b},\delta_{123}\right\}\,.
\end{aligned}
\end{equation}

\subsubsection{Cut $\{z_{3},z_{5}\}$}

On this cut, the diagonal basis is given by
\begin{equation}
\begin{aligned}
    e = \left\{1,\frac{1}{z_{1} z_{2}},\frac{1}{z_{1} z_{4}},\frac{1}{z_{2} z_{4}},\frac{1}{z_{1} z_{2} z_{4}}\right\}\,, \quad h = \left\{\frac{1}{b^{2}},\frac{\delta_{12}}{b},\frac{\delta_{14}}{b},\frac{\delta_{24}}{b},\delta_{124}\right\}\,.
\end{aligned}
\end{equation}

\newpage
\subsection{Massive Pentagon}

\begin{figure}[h!]
\centering
\includegraphics[scale=0.3]{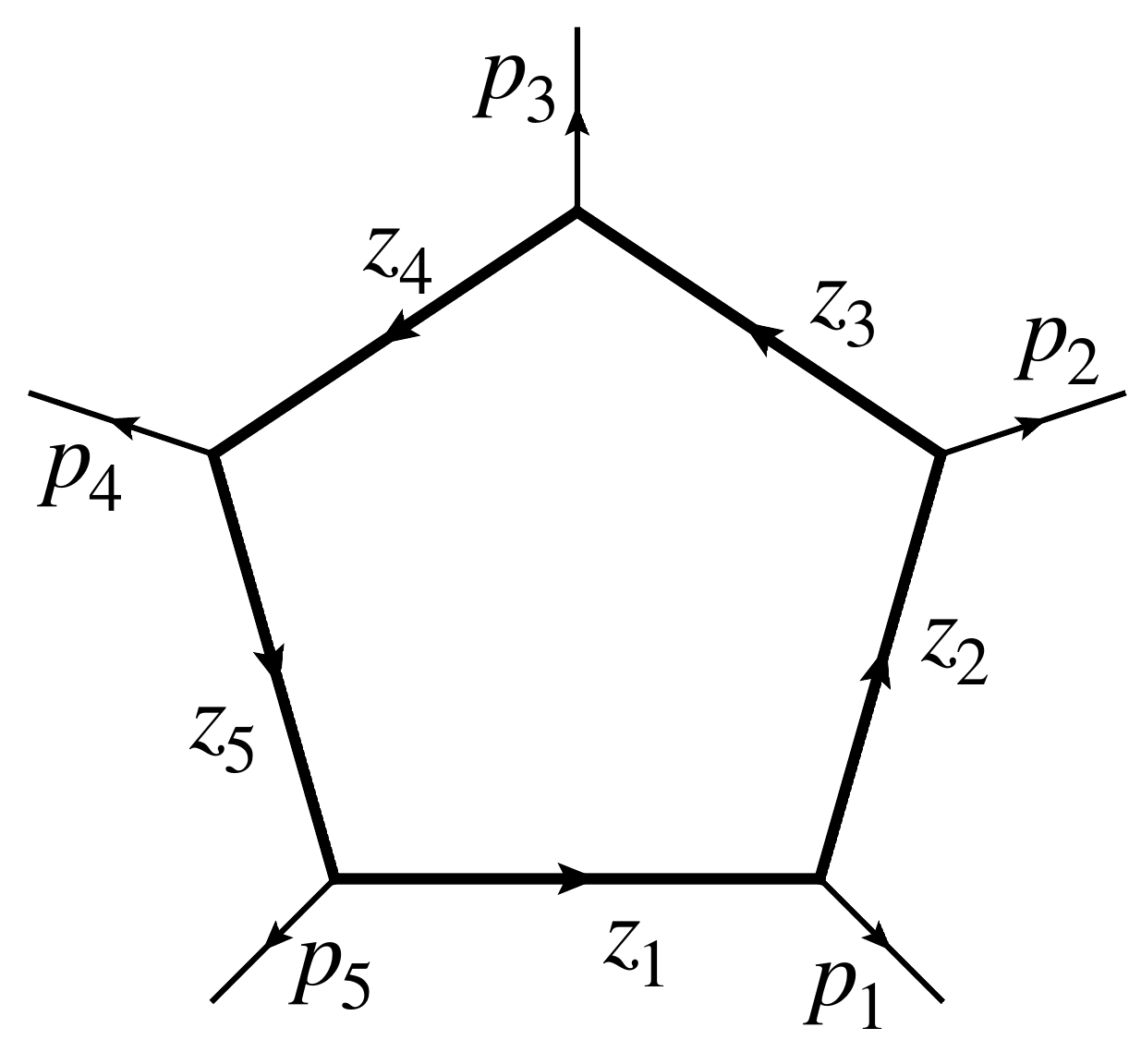}
\caption{One-loop Massive Pentagon with external momenta $p_{1},p_{2},p_{3},p_{4},p_{5}$ and propagators $z_{1},z_{2},z_{3},z_{4},z_{5}$.\label{fig:OLMP}}
\end{figure}

The massive pentagon is shown in Fig. \ref{fig:OLMP}

The propagators are
\begin{equation}
\begin{aligned}
    z_{1} &= \ell^{2}-m_{2}^{2}, \quad z_{2} = (\ell-p_{1})^{2}-m_{2}^{2}, \quad z_{3} = (\ell-p_{1}-p_{2})^{2}-m_{2}^{2}\\
    z_{4} &= (\ell-p_{1}-p_{2}-p_{3})^{2}-m_{2}^{2}, \quad z_{5} = (\ell-p_{1}-p_{2}-p_{3}-p_{4})^{2}-m_{2}^{2}.
\end{aligned}
\end{equation}
The kinematics are
\begin{align}
    p_{i}^{2} = m_{1}^{2}, \quad s_{ij} = (p_{i}+p_{j})^{2}, \quad \sum_{0<i<j<5}s_{ij} = 8m_{1}^{2}
\end{align}
The spanning set of cuts up to symmetry relations is
\begin{align}
    \{z_{5}\}.
\end{align}
\subsubsection{Cut $\{z_{5}\}$}
On this cut, the diagonal basis is given by
\begin{equation}
\begin{aligned}
    e &= \bigg\{1,\frac{1}{z_{1}},\frac{1}{z_{2}},\frac{1}{z_{3}},\frac{1}{z_{4}},\frac{1}{z_{1} z_{2}},\frac{1}{z_{1} z_{3}},\frac{1}{z_{1} z_{4}},\frac{1}{z_{2} z_{3}},\frac{1}{z_{2} z_{4}},\frac{1}{z_{3} z_{4}},\\
    &\qquad\frac{1}{z_{1} z_{2} z_{3}},\frac{1}{z_{1} z_{2} z_{4}},\frac{1}{z_{1} z_{3} z_{4}},\frac{1}{z_{2} z_{3} z_{4}},\frac{1}{z_{1} z_{2} z_{3} z_{4}}\bigg\}\\
    h &= \bigg\{\frac{1}{b^4},\frac{\delta_{1}}{b^3},\frac{\delta_{2}}{b^3},\frac{\delta_{3}}{b^3},\frac{\delta_{4}}{b^3},\frac{\delta_{12}}{b^2},\frac{\delta_{13}}{b^2},\frac{\delta_{14}}{b^2},\frac{\delta_{23}}{b^2},\frac{\delta_{24}}{b^2},\frac{\delta_{34}}{b^2},\frac{\delta_{123}}{b},\frac{\delta_{124}}{b},\frac{\delta_{134}}{b},\frac{\delta_{234}}{b},\delta_{1234}\bigg\}.
\end{aligned}
\end{equation}

\newpage

\subsection{Massless Hexagon}

\begin{figure}[h!]
\centering
\includegraphics[scale=0.45]{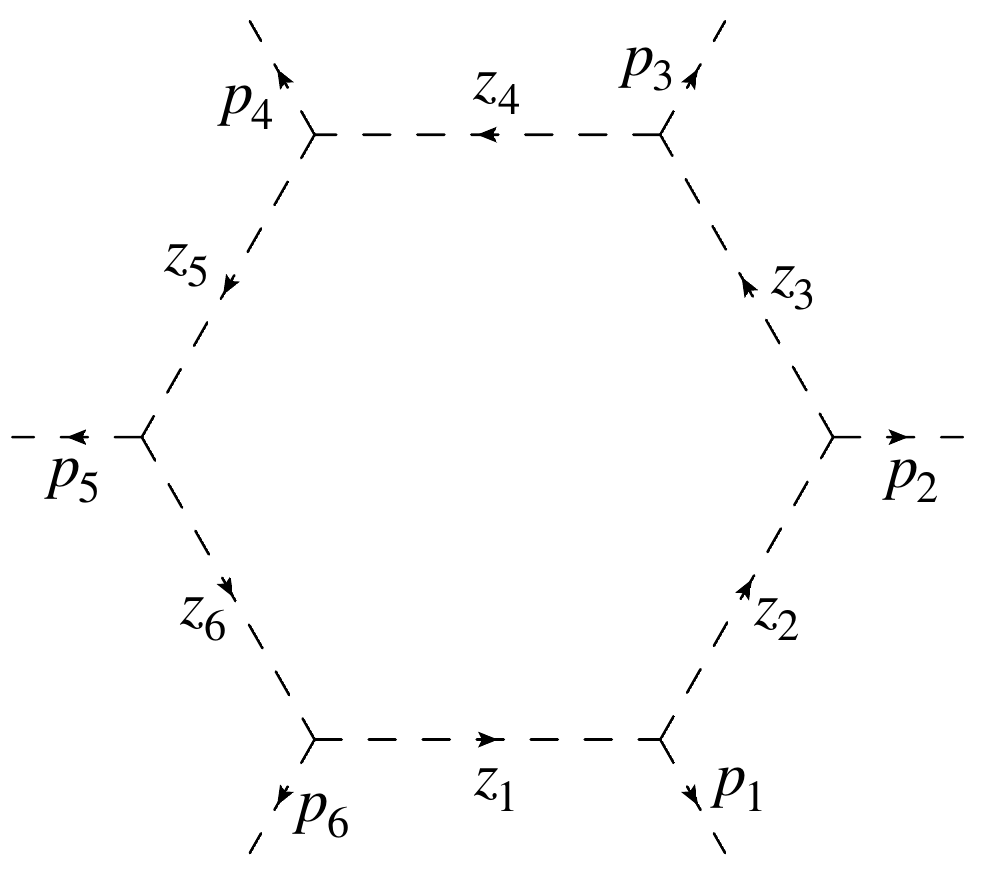}
\caption{One-loop Massless Hexagon with external momenta $p_{1},p_{2},p_{3},p_{4},p_{5},p_{6}$ and propagators $z_{1},z_{2},z_{3},z_{4},z_{5},z_{6}$.\label{fig:OLH}}
\end{figure}

The massless hexagon is shown in Fig. \ref{fig:OLH}

The propagators are
\begin{equation}
\begin{aligned}
    z_{1} &= \ell^{2}\,, \quad z_{2} = (\ell-p_{1})^{2}\,, \quad z_{3} = (\ell-p_{1}-p_{2})^{2}\,,\quad z_{4} = (\ell-p_{1}-p_{2}-p_{3})^{2}\,,\\
    z_{5} &= (\ell-p_{1}-p_{2}-p_{3}-p_{4})^{2}\,, \quad z_{6} = (\ell-p_{1}-p_{2}-p_{3}-p_{4}-p_{5})^{2}\,.
\end{aligned}
\end{equation}
The kinematics are
\begin{align}
    p_{i}^{2} = 0\,, \quad s_{ij} = (p_{i}+p_{j})^{2}\,, \quad \sum_{0<i<j<6}s_{ij} = 0\,.
\end{align}
The spanning set of cuts up to symmetry relations is
\begin{align}
    \{z_{5},z_{6}\}\,,\quad\{z_{4},z_{6}\}\,,\quad\{z_{3},z_{6}\}\,.
\end{align}
It is important to reiterate here that we are considering the external momenta to be $d$-dimensional, and therefore the hexagon appears in the basis, unlike the $4$-dimensional case.

\subsubsection{Cut $\{z_{5},z_{6}\}$}

On this cut, the diagonal basis is given by
\begin{equation}
    \begin{aligned}
        e &= \left\{\frac{1}{z_{1} z_{2}},\frac{1}{z_{1} z_{3}},\frac{1}{z_{1} z_{4}},\frac{1}{z_{2} z_{3}},\frac{1}{z_{2} z_{4}},\frac{1}{z_{3} z_{4}},\frac{1}{z_{1} z_{2} z_{3}},\frac{1}{z_{1} z_{2} z_{4}},\frac{1}{z_{1} z_{3} z_{4}},\frac{1}{z_{2} z_{3} z_{4}},\frac{1}{z_{1} z_{2} z_{3} z_{4}}\right\}\,,\\
        h &= \left\{\frac{\delta_{12}}{b^2},\frac{\delta_{13}}{b^2},\frac{\delta_{14}}{b^2},\frac{\delta_{23}}{b^2},\frac{\delta_{24}}{b^2},\frac{\delta_{34}}{b^2},\frac{\delta_{123}}{b},\frac{\delta_{124}}{b},\frac{\delta_{134}}{b},\frac{\delta_{234}}{b},\delta_{1234}\right\}\,.
    \end{aligned}
\end{equation}

\subsubsection{Cut $\{z_{4},z_{6}\}$}

On this cut, the diagonal basis is given by
\begin{equation}
    \begin{aligned}
        e &= \left\{1,\frac{1}{b z_{2}},\frac{1}{z_{1} z_{2}},\frac{1}{z_{1} z_{3}},\frac{1}{z_{1} z_{5}},\frac{1}{z_{2} z_{3}},\frac{1}{z_{2} z_{5}},\frac{1}{z_{3} z_{5}},\frac{1}{z_{1} z_{2} z_{3}},\frac{1}{z_{1} z_{2} z_{5}},\frac{1}{z_{1} z_{3} z_{5}},\frac{1}{z_{2} z_{3} z_{5}},\frac{1}{z_{1} z_{2} z_{3} z_{5}}\right\}\,,\\
        h &= \left\{\frac{1}{b^3},\frac{\delta_{2}}{b^3},\frac{\delta_{12}}{b^2},\frac{\delta_{13}}{b^2},\frac{\delta_{15}}{b^2},\frac{\delta_{23}}{b^2},\frac{\delta_{25}}{b^2},\frac{\delta_{35}}{b^2},\frac{\delta_{123}}{b},\frac{\delta_{125}}{b},\frac{\delta_{135}}{b},\frac{\delta_{235}}{b},\delta_{1235}\right\}\,.
    \end{aligned}
\end{equation}

\subsubsection{Cut $\{z_{3},z_{6}\}$}

On this cut, the diagonal basis is given by
\begin{equation}
    \begin{aligned}
        e &= \left\{1,\frac{1}{z_{1} z_{2}},\frac{1}{z_{1} z_{4}},\frac{1}{z_{1} z_{5}},\frac{1}{z_{2} z_{4}},\frac{1}{z_{2} z_{5}},\frac{1}{z_{4} z_{5}},\frac{1}{z_{1} z_{2} z_{4}},\frac{1}{z_{1} z_{2} z_{5}},\frac{1}{z_{1} z_{4} z_{5}},\frac{1}{z_{2} z_{4} z_{5}},\frac{1}{z_{1} z_{2} z_{4} z_{5}}\right\}\,,\\
        h &= \left\{\frac{1}{b^3},\frac{\delta_{12}}{b^2},\frac{\delta_{14}}{b^2},\frac{\delta_{15}}{b^2},\frac{\delta_{24}}{b^2},\frac{\delta_{25}}{b^2},\frac{\delta_{45}}{b^2},\frac{\delta_{124}}{b},\frac{\delta_{125}}{b},\frac{\delta_{145}}{b},\frac{\delta_{245}}{b},\delta_{1245}\right\}\,.
    \end{aligned}
\end{equation}

\subsection{Massive Hexagon}

\begin{figure}[h!]
\centering
\includegraphics[scale=0.45]{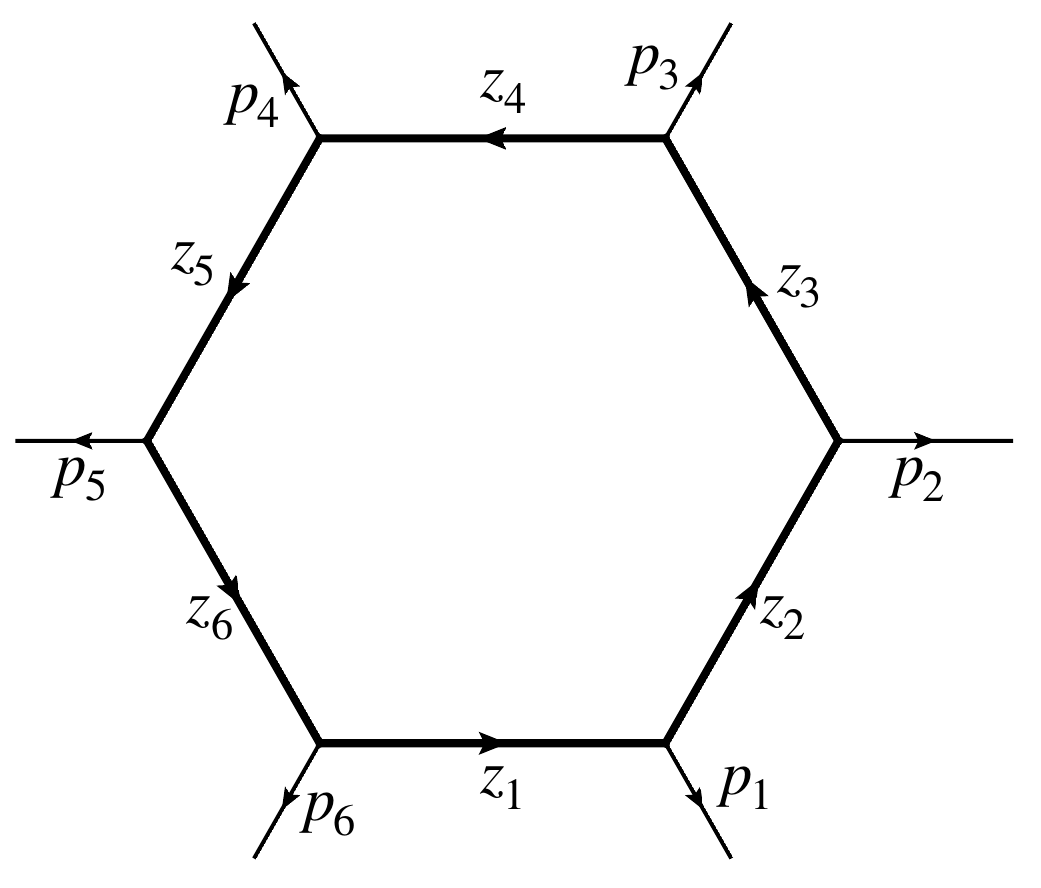}
\caption{One-loop Massive Hexagon with external momenta $p_{1},p_{2},p_{3},p_{4},p_{5},p_{6}$ and propagators $z_{1},z_{2},z_{3},z_{4},z_{5},z_{6}$.\label{fig:OLMH}}
\end{figure}

The massive hexagon is shown in Fig. \ref{fig:OLMH}

The propagators are
\begin{equation}
\begin{aligned}
    z_{1} &= \ell^{2}-m_{2}^{2}\,, \quad z_{2} = (\ell-p_{1})^{2}-m_{2}^{2}\,, \quad z_{3} = (\ell-p_{1}-p_{2})^{2}-m_{2}^{2}\,,\\
    z_{4} &= (\ell-p_{1}-p_{2}-p_{3})^{2}-m_{2}^{2}\,, \quad z_{5} = (\ell-p_{1}-p_{2}-p_{3}-p_{4})^{2}-m_{2}^{2}\,,\\
    z_{6} &= (\ell-p_{1}-p_{2}-p_{3}-p_{4}-p_{5})^{2}-m_{2}^{2}\,.
\end{aligned}
\end{equation}
The kinematics are
\begin{align}
    p_{i}^{2} = m_{1}^{2}\,, \quad s_{ij} = (p_{i}+p_{j})^{2}\,, \quad \sum_{0<i<j<6}s_{ij} = 15m_{1}^{2}\,.
\end{align}
The spanning set of cuts up to symmetry relations is
\begin{align}
    \{z_{6}\}\,.
\end{align}
\subsubsection{Cut $\{z_{6}\}$}
On this cut, the diagonal basis is given by
\begin{equation}
    \begin{aligned}
        e &= \bigg\{1,\frac{1}{z_{1}},\frac{1}{z_{2}},\frac{1}{z_{3}},\frac{1}{z_{4}},\frac{1}{z_{5}},\frac{1}{z_{1} z_{2}},\frac{1}{z_{1} z_{3}},\frac{1}{z_{1} z_{4}},\frac{1}{z_{1} z_{5}},\frac{1}{z_{2} z_{3}},\frac{1}{z_{2} z_{4}},\frac{1}{z_{2} z_{5}},\frac{1}{z_{3} z_{4}},\frac{1}{z_{3} z_{5}},\frac{1}{z_{4} z_{5}},\\
        &\qquad\frac{1}{z_{1} z_{2} z_{3}},\frac{1}{z_{1} z_{2} z_{4}},\frac{1}{z_{1} z_{2} z_{5}},\frac{1}{z_{1} z_{3} z_{4}},\frac{1}{z_{1} z_{3} z_{5}},\frac{1}{z_{1} z_{4} z_{5}},\frac{1}{z_{2} z_{3} z_{4}},\frac{1}{z_{2} z_{3} z_{5}},\frac{1}{z_{2} z_{4} z_{5}},\frac{1}{z_{3} z_{4} z_{5}},\\
        &\qquad\frac{1}{z_{1} z_{2} z_{3} z_{4}},\frac{1}{z_{1} z_{2} z_{3} z_{5}},\frac{1}{z_{1} z_{2} z_{4} z_{5}},\frac{1}{z_{1} z_{3} z_{4} z_{5}},\frac{1}{z_{2} z_{3} z_{4} z_{5}},\frac{1}{z_{1} z_{2} z_{3} z_{4} z_{5}}\bigg\}\,.\\
        h &= \bigg\{\frac{1}{b^5},\frac{\delta_{1}}{b^4},\frac{\delta_{2}}{b^4},\frac{\delta_{3}}{b^4},\frac{\delta_{4}}{b^4},\frac{\delta_{5}}{b^4},\frac{\delta_{12}}{b^3},\frac{\delta_{13}}{b^3},\frac{\delta_{14}}{b^3},\frac{\delta_{15}}{b^3},\frac{\delta_{23}}{b^3},\frac{\delta_{24}}{b^3},\frac{\delta_{25}}{b^3},\frac{\delta_{34}}{b^3},\frac{\delta_{35}}{b^3},\frac{\delta_{45}}{b^3},\\
        &\qquad\frac{\delta_{123}}{b^2},\frac{\delta_{124}}{b^2},\frac{\delta_{125}}{b^2},\frac{\delta_{134}}{b^2},\frac{\delta_{135}}{b^2},\frac{\delta_{145}}{b^2},\frac{\delta_{234}}{b^2},\frac{\delta_{235}}{b^2},\frac{\delta_{245}}{b^2},\frac{\delta_{345}}{b^2},\\
        &\qquad\frac{\delta_{1234}}{b},\frac{\delta_{1235}}{b},\frac{\delta_{1245}}{b},\frac{\delta_{1345}}{b},\frac{\delta_{2345}}{b},\delta_{12345}\bigg\}\,.
    \end{aligned}
\end{equation}

\subsection{Two-loop Example: Elliptic Sunrise}

\begin{figure}[h!]
\centering
\includegraphics[scale=0.4]{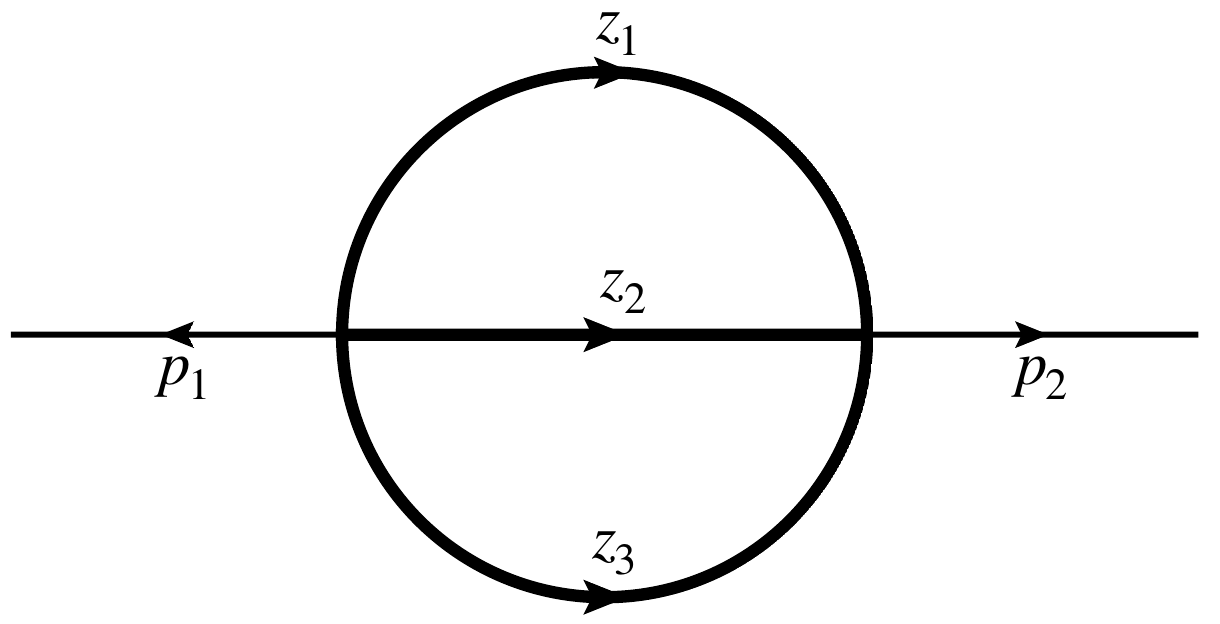}
\caption{Two-loop Massive Sunrise with external momenta $p_{1},p_{2}$ and propagators $z_{1},z_{2},z_{3}$.\label{fig:sunrise}}
\end{figure}

The elliptic sunrise is shown in Fig. \ref{fig:sunrise}

The propagators are
\begin{equation}
\begin{aligned}
    z_{1} = \ell_{2}^2-m_2^2, \quad z_{2} = (\ell_{1}-\ell_{2})^2-m_2^2,\quad z_{3}=(\ell_{1}+p_{1})^2-m_2^2,\quad z_{4}=\ell_{1}^2,\quad z_{5}=(\ell_{2}+p_{1})^2\,.
\end{aligned}
\end{equation}
The kinematics are $p_{i}^{2}=m_{1}^{2}$. Our basis choice is
\begin{equation}
    \begin{aligned}
        e &= \left\{\frac{1}{z_{1} z_{2}},\frac{1}{z_{1} z_{3}},\frac{1}{z_{2} z_{3}},\frac{1}{z_{1} z_{2}^2 z_{3}},\frac{1}{z_{1} z_{2} z_{3}^2},\frac{1}{z_{1} z_{2} z_{3}}\right\}\\
        h &= \left\{\frac{\delta_{12}}{b^{2}},\frac{\delta_{13}}{b^{2}},\frac{\delta_{23}}{b^{2}},\frac{\partial_{z_2}\delta_{123}}{b},\frac{\partial_{z_3}\delta_{123}}{b},\frac{\delta_{123}}{b}\right\}\,.
    \end{aligned}
\end{equation}
where we use the ``derivative" of the delta form as defined in eq. (\ref{eq:delta_dervative}).
The $\mathbf{C}$-matrix is block diagonal and is given by
\begin{equation}
    \mathbf{C}=\left(
\begin{array}{cccccc}
 \frac{64}{(d-2)^3} & 0 & 0 & 0 & 0 & 0 \\
 0 & \frac{64}{(d-2)^3} & 0 & 0 & 0 & 0 \\
 0 & 0 & \frac{64}{(d-2)^3} & 0 & 0 & 0 \\
 0 & 0 & 0 & -\frac{2}{m^2} & 0 & \frac{8}{3 d-10} \\
 0 & 0 & 0 & 0 & -\frac{4}{m^2} & \frac{8}{3 d-10} \\
 0 & 0 & 0 & -\frac{8}{3 d-8} & -\frac{8}{3 d-8} & \frac{16 \left(3 m^2+s\right)}{(3 d-10) (3 d-8)} \\
\end{array}
\right)\,.
\end{equation}

\section{Conclusions}\label{sec:conclusions}

In this paper we presented two advancements for the computation of intersection numbers for twists with quadratic polynomials. The first of these was a new prescription that produces duals orthogonal to Laporta bases for relative twisted cohomology groups, resulting in diagonal $\mathbf{C}$-matrices. We tested the procedure with a number of one-loop Feynman integrals and found that it works consistently in the fully massive case. If the diagrams were massless our prescription also worked but with some minor modifications. Beyond one-loop, we also tested our result for the elliptic sunrise, and produced a block diagonal $\mathbf{C}$-matrix, with each block representing a (sub)-sector, by making use of ``derivatives" of the delta forms.

The second result was the introduction of a new, closed formula for intersection numbers between quadratic polynomials raised to arbitrary powers. We believe this result to be significant as it is the first such proposition beyond $\dd\log$ forms. We verified this formula to hold with many arbitrarily chosen quadratic polynomials, as well as multiple $1$-loop Feynman Integrals presented in the work. The closed formula, combined with the diagonal basis prescription, allowed for a very efficient computation of $\mathbf{C}$-matrices, by requiring no explicit algorithmic computations of intersection numbers.

Our results presented in this work open many possibilities for further exploration. An important first question would be why the diagonal basis prescription requires modifications when fewer masses are present in the diagrams. In a different direction, one could attempt to generalise the algorithm to two-loop diagrams, beyond the elliptic sunrise. For these cases, where irreducible scalar products play a role and there exist multiple master integrals in each sector, one might hope for a similar prescription to produce block-diagonal $\mathbf{C}$-matrices for more general classes of integrals.

Our new closed formula also presents new interesting questions. Firstly, it would be of great importance to prove this result. Beyond quadratics, one could also ask whether intersection numbers between cubic polynomials or beyond follow similar patterns, as this would represent Feynman integrals beyond one-loop. It would not be unreasonable to expect the multivariate discriminant to appear in such cases also, although in practice from a computational standpoint no closed formula is expected for multivariate discriminants of higher order polynomials.

Finally, our closed formula is specific to twists with a single polynomial factor. Extending this to even just two factors would allow for the computation of a much larger class of intersection numbers, including the projections necessary to complete IBP reductions once the $\mathbf{C}$-matrix has been computed. This could pave the way for a closed formulae for complete IBP reductions at one-loop and perhaps beyond. 

Indeed, in this work we have uncovered mathematical structures in intersection numbers beyond $1$- and $\dd \log$ forms, venturing beyond where many of their properties are well known and studied. Whilst just a small start, a deeper understanding of the characteristics of more general types of intersection numbers could prove instrumental in our future understanding of the Feynman integral vector space and beyond.

\subsection*{Acknowledgements}

We wish to thank Pierpaolo Mastrolia, Wojciech Flieger, Manoj Mandal, Vsevolod Chestnov, Federico Gasparotto, Mathieu Giroux, Giacomo Brunello and Sebastian Mizera for ideas, comments and feedback. We would additionally like to thank Andrzej Pokraka and Hjalte Frellesvig for invaluable discussions relating to all parts of the work. G.C. thanks Saiei-Jaeyeong Matsubara-Heo, Julian Miczajka and Francesco Calisto for stimulating conversations related to multivariate discriminants, as well as Johannes Henn and the Max Planck institute for Physics for hosting him during the development of this project.

\newpage
\appendix

\section{Relation Between the Closed Formula and Multivariate Discriminant}\label{app:multivariate_discriminant}

The multivariate discriminant \cite{Gelfand1994DiscriminantsRA,citeulike:14125041,DANDREA200159} for a homogeneous polynomial $p(\mathbf{z})$ in $n>1$ variables $\mathbf{z}$ is a polynomial in the coefficients of $p$. Specifically, $\mathrm{Disc}_{\mathbf{z}}(p)=0$ if and only if there exists a non trivial solution to the set of equations:
\begin{equation}\label{eq:multivardiscriminant_def}
\begin{aligned}
    \frac{\partial p}{\partial z_1}&=0\\
    &\vdots\\
    \frac{\partial p}{\partial z_n}&=0\
\end{aligned}
\end{equation}
for the variables $\mathbf{z}$. This fixes the multivariate discriminant uniquely up to a numerical prefactor. The ``non trivial'' is necessary because for homogeneous polynomials there will always be the trivial solution $\mathbf{z}=\mathbf{0}$ to eq. (\ref{eq:multivardiscriminant_def}), which we exclude. For an inhomogeneous polynomial $p$ in $n$ variables, the polynomial is first homogenised to obtain an $n+1$ variable polynomial $p_h$. The above definition is then applied to $p_h$. Namely, we have
\begin{equation}
    \mathrm{Disc}_{\mathbf{z}}(p):=\mathrm{Disc}_{\mathbf{y}}(p_h)\,,\qquad \mathbf{y}=\{\mathbf{z},z_{n+1}\}\,.
\end{equation}
Let us now specify to $n$ variable quadratic polynomials, such as $b(\mathbf{z})$ in Section (\ref{subsec:closed_formula}). To show that $\det\left(\mathbf{H}(b_h)\right)=\mathrm{Disc}_{\mathbf{z}}(b)$ we note that if $b$ is quadratic, then $b_h$ is purely quadratic (only contains quadratic terms). Thus, $b_h$ can be written as
\begin{equation}
    b_h=\mathbf{y}^{T}\,A\,\mathbf{y}\,.
\end{equation}
where $A$ is a matrix of coefficients. It is easy show that
\begin{equation}
    \frac{\partial b_h}{\partial \mathbf{y}}=(A^{T}+A)\,\mathbf{y}\,,\qquad\mathbf{H}(b_h)=A^{T}+A\,.
\end{equation}
Thus, eq. (\ref{eq:multivardiscriminant_def}) reduces to
\begin{equation}
    \mathbf{H}(b_h)\,\mathbf{y}=0\,.
\end{equation}
For any solutions other than $\mathbf{y}=\mathbf{0}$ to exist, we must necessarily have $\det\mathbf{H}(b_h)=0$. Thus, we conclude that $\det\mathbf{H}(b_h)$ (up to a numerical prefactor) must be the multivariate discriminant of $b_h$ and equivalently of $b$.

\bibliographystyle{JHEP}
\bibliography{biblio}
\end{fmffile}
\end{document}